# Adaptability and the Pivot Penalty in Science and Technology

Ryan Hill[1,2¶], Yian Yin[1,3,4,5¶], Carolyn Stein[6,7], Xizhao Wang[2], Dashun Wang[1,2,3,4*], Benjamin F. Jones[1,2,3,8*]

[1]Center for Science of Science and Innovation, Northwestern University, Evanston IL
[2]Kellogg School of Management, Northwestern University, Evanston IL
[3]Northwestern Institute on Complex Systems, Northwestern University, Evanston IL
[4]McCormick School of Engineering, Northwestern University, Evanston IL
[5]Department of Information Science, Cornell University, Ithaca NY
[6]Haas School of Business, University of California – Berkeley, Berkeley CA
[7]Department of Economics, University of California – Berkeley, Berkeley CA
[8]National Bureau of Economic Research, Cambridge MA
[¶]These authors contributed equally to this work.
[*]Correspondence to: dashun.wang@northwestern.edu, bjones@kellogg.northwestern.edu.

August 2024

**Abstract**

**Scientists and inventors set the direction of their work amidst an evolving landscape of questions, opportunities, and challenges. This paper introduces a measurement framework to quantify how far researchers move from their existing research when producing new works. We apply this framework to millions of scientific publications and patents and uncover a pervasive "pivot penalty," where the impact of new research steeply declines the further a researcher moves from their prior work. The pivot penalty applies nearly universally across scientific publishing and patenting and has been growing in magnitude over the past five decades. While creativity frameworks suggest a benefit to exploratory search by researchers and often emphasize outsider advantages in driving breakthroughs, we find little evidence for such an advantage. The pivot penalty is consistent with increasingly narrow specializations of researchers, and when researchers undertake large pivots, a signature of their work is weak engagement with established mixtures of prior knowledge. Unexpected shocks to the research landscape, which may push researchers away from existing areas or pull them into new ones, further demonstrate substantial pivot penalties. COVID-19 provides a high-scale case study, where many researchers engaged the pandemic, yet the pivot penalty remains severe. The pivot penalty generalizes across fields, career stage, productivity, collaboration, and funding contexts, highlighting both the breadth and depth of the adaptive challenge. Overall, the findings point to large and increasing challenges in adapting to new opportunities and threats. The results have implications for individual researchers, research organizations, science policy, and the capacity of science and society as a whole to confront emergent demands.**

## Introduction

Science has been described as an endless frontier [1-4], engaging an evolving array of questions, opportunities, and challenges [5, 6]. New areas emerge, from synthetic biology to climate change to the COVID-19 pandemic, and both researchers and research organizations must consider adapting their research portfolios to address emergent opportunities and demands [7-11]. Adaptability to threats and opportunities is thus critical to scientific and technological progress [1-3], and adaptive success or failure can underpin the relative progress or collapse of organizations, economic regions, and societies [1-4, 12, 13].

The adaptability of research streams hinges on researchers themselves, who must regularly consider the direction of their work and their potential to engage new areas. Yet while researchers face consequential choices across large or small changes in their research directions, the degree to which research directions are adaptable depends on fundamental tradeoffs and unknowns. On the one hand, shifts in research may be difficult [14]. The specialization of expertise [15, 16], the design of funding systems [17, 18], and the nature of research incentives, culture, and communities [19-22] may all limit the capacity of a given individual to respond effectively to changing opportunities and demands [23-27]. On the other hand, the value of novelty [28-30] and exploration [31-33] in creative search suggests that reaching further from one's usual research area might be especially fruitful [14, 34, 35], and new entrants or "outsiders" to a given area are sometimes thought to be especially capable of transformative ideas [22, 36]. Indeed, a researcher who continues to exploit an existing direction may face diminishing returns while missing opportunities afforded in other areas [31, 37]. Exploring new areas might then be risky but also more likely to produce high impact insights.

Here we study the adaptability of scientists and inventors, examining outcomes when researchers work in areas nearer or further from their existing research portfolio. We introduce a measurement framework for research "pivots" and then study adaptability in both general and specific settings. We first apply the measurement framework at high scale across scientific and technological domains, studying millions of scientific articles indexed by Dimensions from 1970-2020 and U.S. patents granted from 1985-2020 (see SM 1.1-2). The core finding is a substantial "pivot penalty," where the impact of new research steeply declines the further a researcher moves from their prior work. This pivot penalty appears within individual researchers, across wide-ranging fields of inquiry, and has been steepening over time. We then evaluate the pivot penalty in light of canonical conceptual frameworks and investigate potential mechanisms, drawing on ideas of reputation and audience [27, 38-40] as well as creativity frameworks in the production of new ideas [15, 28, 31]. Finally, we turn to case studies of substantial interest to science and where exogenous events can elicit research pivots. We first study "push" events, where existing knowledge is revealed to be incorrect or unreliable, pushing researchers away from prior research streams. We then study a "pull" event – the COVID-19 pandemic – which acted to pull researchers into an important new research area. We find that despite the wide-ranging nature of these events, researchers consistently pivot to an unusually large degree after these events and that the pivot penalty persists in each case. The pandemic further allows us to examine a consequential, society-scale event and the capacity of science as a whole to address novel research demands. The paper concludes with discussion of implications of these findings for researchers, research organizations, and science policy.

**Measurement Framework**

To quantify pivots for researchers, we calculate a cosine-similarity metric (Fig. 1A) that measures the extent to which a given new work departs from a researcher's prior body of work. Specifically, in the sciences, for an author $i$ and a focal paper $j$, we calculate a vector $R_i^j$, representing the distribution of journals referenced by $j$. Similarly, we count the frequency in which different journals are referenced in the union of $i$'s prior work, defining a vector $R_i$. An individual's works include any paper where the individual is a listed author. The pivot measure, $\Phi_i^j$, is then defined as 1 minus the cosine of these two vectors:

$$\Phi_i^j = 1 - \frac{R_i^j \cdot R_i}{\left\|R_i^j\right\| \left\|R_i\right\|} \tag{1}$$

The measure $\Phi_i^j$ thus takes the value 0 ("zero pivot") if the focal paper draws on the exact same distribution of journals as the author's prior work and takes the value 1 ("full pivot") if the focal paper draws entirely on a novel set of journals. The measure featured in the main text calculates pivoting in the focal paper compared to the prior three years of the author's work. We also calculate our measure by using all prior work of a given author, arriving at similar conclusions (see SM and Fig. S1). In the patent context, where journal information is not available, we use technological field codes to measure pivots. Specifically, we use the distribution of Cooperative Patent Classification (CPC) technology field codes among a patent's cited references to build the reference vectors and cosine similarity metric in (1). These technology codes are hierarchical, providing alternative levels of granularity in defining technology areas. Our main analyses use the detailed level-4 technological classification (comprising 9,987 distinct technology areas), and we further examine all possible classification levels in the supplementary material.

Fig. 1 shows the distribution of pivoting behavior using this measurement framework and focusing on the year 2020. Overall, we see wide dispersion of pivoting in both the science and patenting contexts, suggesting that pivoting is prevalent for both scientists and inventors, and the size of pivots has high variance (Fig. 1B-C). Anticipating our case study of COVID-19 research, we observe a sharp difference in pivot size comparing COVID-19 and non-COVID-19 related research, where scientists who engaged COVID-19 exhibit unusually large pivots. Whereas non-COVID papers in 2020 present a median of $\overline{\Phi} = 0.60$, COVID-19 papers present a substantially larger median pivot size of $\overline{\Phi} = 0.82$ ($p<.0001$).

The highly variable nature of the pivot size is especially prominent in patenting, where we observe a bimodal distribution (Fig. 1C) with weight at both extremes, showing a tendency for both small pivots and large jumps. Given the hierarchical structure of patent technology codes, we can further examine pivoting from the broadest level-1 classification level (9 sections) to the most detailed level-5 classification (210,347 subgroups). Intuitively, the pivot distribution for inventors shifts leftward when using broader technology categories (Fig. S2). In other words, inventors pivot less from their broadest technology areas (the section or section-class level). Similarly, for papers, we can use a coarser, field-level coding instead of journals to measure pivots and observe a leftward shift in the pivot size distribution (Fig. S3). Yet, regardless of the categories we use, pivot behavior remains dispersed.

**The Pivot Penalty**

As scientists and inventors shift from their earlier research, a central question is how impactful their work becomes. We first consider 26 million papers published from 1970 to 2015 across 154 fields. To quantify impact, we calculate the paper-level hit rate, a binary indicator for whether a given work was in the upper 5% of citations received within its field and publication year [41]. Fig. 2A reveals a striking fact: looking at all of the science, works with larger average pivots are systematically associated with lower impact. Indeed, we observe a large and monotonic decrease in the average hit rate as the pivot size rises. The lowest-pivot work is high impact 7.4 percent of the time, 48% higher than the baseline rate, whereas the highest-pivot work is high impact only 2.2 percent of the time, a 56% reduction from the baseline. Fig. 2B normalizes impact within individual researchers using regressions with individual fixed effects (see SM S2.5 for methods), showing a pivot penalty that is both substantial and somewhat less steep than in the raw data. Within a given researcher's portfolio, the lowest-pivot work is 2.1 percentage points ($p<.001$) more likely to be high impact than that researcher's other work, while their highest-pivot work is 1.8 percentage points less likely ($p<.001$) to be high impact, again showing large deviations from the 5 percent baseline. A range of robustness tests, including those measuring citation impact in a continuous manner or over different time horizons, produces similar findings (see SI S2.2).

We next consider 1.8 million patents granted from 1980-2015 across 127 technology classes and similarly calculate the patent-level hit rate based on being in the upper 5% of citations received within the patent's technology classification and application year. We find again a monotonic decrease in impact as pivot size increases (Fig. 2C). The lowest pivot patents are high impact 8.0 percent of the time, 60% higher than the baseline rate, while the highest-pivot patents are high impact only 3.8 percent of the time, a 24% reduction from the baseline. This decline in impact with larger pivots is robust to measuring inventor pivots at any technology-classification level, from the broadest to the narrowest (Fig. S4). Fig. 2D further normalizes impact within individual inventors and continues to show the pivot penalty.

Quantifying this "pivot penalty" over time, we find the relationship between pivot size and impact in science has become increasingly negative over the past five decades. The steepening of the pivot penalty appears in the raw data (Fig. 2E) and when looking within individual researchers (Fig. S5). Furthermore, these findings generalize widely not just across time but also across scientific fields. Studying separately each of the 154 subfields, we find that the negative relationship between impact and pivot size holds for 93% of fields, and the increasing severity of the pivot penalty over time occurs in 88% of all scientific fields (Table S1). Turning to patenting, we again observe an increasingly steep pivot penalty with time (Fig. 2F). Studying separately 127 level-2 technology classes, we find that the negative relationship between impact and pivot size holds in 91% of classes, with the severity of the pivot penalty growing over time in 76% of patent classes (Table S2). This steepening pivot penalty among inventors is also seen when using broader or narrower technological classifications (Fig. S6). Earlier years for patenting show flatter, less monotonic relationships in the raw data (Fig. 2F) and within inventors (Fig. S7).

We further find that the pivot penalty is robust to many alternative measures and sample restrictions, and we also perform deeper dives into high pivot cases and outlier fields (see SI section S3). Robustness tests include alternative time windows to determine citation impact (Fig S8); alternative citation impact measures (Fig. S9); sample restrictions to papers with larger reference counts (Table S3); redoing pivot size computation based on referenced papers' coarser field coding as opposed to their journals (Fig. S3); and hand checks on high-pivot researchers (Section S3). These analyses all support the robustness of our main results.

In examining outcomes, one can also look beyond citation impact. For papers, we further measure whether a paper is referenced in a future patent [3, 42], indicating the use of the idea beyond science. We see a large decline in patent references to high-pivot articles, where the probability of being cited in a patented invention declines by 43% comparing the highest-pivot to the lowest-pivot papers and with some non-monotonicity at low pivot sizes (Fig. S10). We also examine the propensity for preprints to become published and find that high-pivot preprints publish at lower rates, indicating another form of the pivot penalty (Fig. S11). For patents, we consider the invention's market value based on how a company's stock price moves in response to the patent's issuance [43]. We see that the market value of a patented invention declines steeply with pivot size, with the market value declining 29% comparing the highest-pivot to the lowest-pivot patents (Fig. S12). These findings indicate that the pivot penalty also appears when considering practical use and market value, and beyond the citation behavior within a community of researchers.

Altogether, we observe striking empirical regularities that generalize across science and technology. Despite the distinct nature of scientific articles and patents, the different institutional contexts in which they are produced, the wide range of research fields, and the alternative outcome measures, these spheres present remarkable commonalities: For both scientists and inventors, greater pivots present large impact penalties, and increasingly so with time.

**Conceptual Frameworks and Potential Mechanisms**

The impact advantage to narrowness in the ambit of research suggests substantial difficulties for researchers in entering new areas. It further heightens concerns in innovation communities that research with wide reach or novel orientations is difficult [10, 14-16, 19]. Entering new areas may be challenging as a matter of reception, where a scholar has difficulty penetrating new audiences, and it may be challenging as a matter of idea generation, where scholars face difficulties generating valuable ideas outside their key areas of competency. To further inform the nature of the pivot penalty, we next examine the pivot penalty in view of both reputational perspectives and idea generation frameworks.

An established reputation in a local research community may provide impact advantages within that community and a relative disadvantage outside it [38]. For example, the "Matthew Effect" [38, 39] suggests advantages of established eminence within a community, while "typecasting" [27, 40] may undermine receptions when entering new areas. These and other reputational considerations suggest that the pivot penalty may emerge because researchers move beyond their usual field or audience. To test these considerations, we first examine pivots holding the researcher's field or local audience fixed. Specifically, we examine what happens when a given researcher publishes multiple papers with different pivot sizes but in the same time frame and field, and even in the same exact journal (Table S4). We find that the pivot penalty is approximately 28% less steep when an individual is publishing in the same journal, an attenuation consistent with a weakening of reputational forces when looking within a common audience. Yet the large majority of the relationship remains. The pivot penalty thus remains strong for a given researcher when publishing in a consistent field or before a consistent, local readership. A related approach considers impact within a given, distant audience. Recalling the findings for market value (Fig. S12) and patented applications (Fig. S10), the pivot penalty also appears when examining how inventors draw on science or how investors value inventions. These evaluations are made by individuals who are far away from the focal researcher. In sum, the pivot penalty appears not simply as a matter of movement across fields or from a local audience to a distant audience. Rather it appears for a researcher within a given field or journal, and it appears within distant communities tuned to practical use and market returns.

Reputational considerations may be further informed by considering career stage. Specifically, younger researchers, with less formed reputations, may see less advantage (the Matthew Effect) from staying in a given area or less penalty (typecasting) from venturing outside it [40, 44]. Studying career stage, we find some evidence that the pivot penalty appears slightly stronger with advancing career stages, consistent with these reputational frameworks. Yet the pivot penalty remains strong regardless of career stage, including very early in the career (Table S5). The findings continue to suggest adaptive challenges, beyond the force of established reputations, when entering new research terrain.

Turning to idea generation frameworks, a canonical perspective emphasizes an "explore vs. exploit" tradeoff in creative search. Here, exploitation involves lower-risk but potentially lower-return search around the edges of one's current focus, while exploration involves higher-risk but potentially higher-return departures into more distant areas [31, 36, 37]. Related views suggest an advantage of outsiders in bringing novel perspectives and driving breakthroughs [34, 35, 45]. Our analyses have looked at upper-tail outcomes, but it is possible that the value of large pivots lies in even rarer and more extreme positive outcomes. Surprisingly, however, we find that high-pivot research receives lower citations across the entire citation distribution (Fig. S13A). In the very upper tail, such as in the upper 1% or upper 0.1% of citation impact, high-pivot work is even more heavily underrepresented compared to low-pivot work (Fig. S13B). Rather than suggesting a tradeoff between risk and reward in exploratory search, or outsider advantages, these findings continue to suggest a more fundamental difficulty of venturing into new areas.

Alternative idea generation frameworks emphasize the value of specialized expertise. These frameworks link creative advantages less to outsider ideas and more to the accumulated facts, theories, and methods built in an area by prior scholars [45, 46]. The emphasis on expertise and the value of prior knowledge is consistent with Newton's famous statement that "if I have seen further, it is by standing on the shoulders of giants" [47]. Further, the steepening of the pivot penalty with time is consistent with increasingly narrow expertise as science progresses and knowledge deepens [15, 48, 49]. Related, creativity frameworks emphasize that new works can be seen as new combinations of existing material [50-52]. Prior literature has shown that high-impact research is characterized primarily by highly conventional mixtures of prior knowledge while also tending to inject, simultaneously, a small dose of atypical combinations that are unusual in previous research [28, 53]. Following this literature, we further measure the novelty and conventionality of combinations in a given paper and relate these measures to pivot size (see SI S2.3 for methods). We find that high-pivot work is associated with a higher propensity for atypical combinations (Fig. S14A), a feature also reflected in earlier work linking inventors who switch fields to novel technology combinations [14]. In other words, when pivoting, a researcher not only does something new personally but also tends to introduce novel combinations of knowledge to the broader research domain. Yet, at the same time, high-pivot papers show distinctly low conventionality (Fig. S14B), locating a key characteristic that such exploratory work tends to miss: a deep grounding in established mixtures of knowledge. These findings suggest that researchers, as they shift to new areas personally, are equipped for novelty but limited in their relevant or conventional expertise, underscoring the difficulty researchers may face in venturing beyond their specialized knowledge.

**Pivoting in Response to External Events**

The pivot penalty indicates that larger pivots are strongly associated with lower impact. Yet, at the same time, the research landscape itself is constantly shifting, and researchers must weigh opportunities nearer to and further from their current research streams. To further probe pivoting behavior and the pivot penalty, we consider external events that may provoke researchers to pivot. External events can provide quasi-experimental settings and help to establish causal interpretations of the pivot penalty while further informing the tensions in how researchers navigate a shifting research landscape.

We first consider events that may push researchers away from an existing research stream. Specifically, prior research is sometimes revealed as incorrect or unreliable, which may encourage researchers who had been building on that work to move in new directions. Here we focus on paper retractions, which occur at relatively high scale and are of growing interest to the science community [54-56]. Using Retraction Watch and the Dimensions database, we locate 13,456 retractions over the 1975-2019 period. As a treatment group, we consider researchers whose work referenced a retracted paper prior to its retraction (but who were not authors of the retracted study). As a control group, we consider researchers who referenced other papers appearing in the same journal and year as the retracted paper. We further use coarsened exact matching [57] to match treated and control authors by their publication rates prior to the retraction year. We then compare pivots and hit rates between the treatment and control groups, over the four years before and four years after retraction events, in a difference-in-differences design. See Fig. 3A (further methods details in SI S2.6).

We find that pivot sizes increase markedly after a retraction event (Fig. 3B). Consider first the 146,744 treated researchers who referenced a retracted paper at least once prior to its retraction. The mean pivot size for these researchers' works after the retraction increases by 2.5 percentage points ($p<.001$) compared to control researchers' works. We further study a smaller treatment group of 16,413 researchers who referenced a retracted paper multiple times, indicating more intensive use. For this group, pivoting is larger, with mean pivot sizes increasing by 3.7 percentage points ($p<.001$) after the retraction, compared to the control authors (Fig. 3B).

We next examine paper impact. Treated authors experience a 0.4 percentage point decline ($p<.001$) in hit rate after the shock, compared to control authors (Fig. 3B). Among treated authors who drew on the retracted study multiple times, we see not only larger pivots (Fig. 3B, left) but also a larger 0.7 percentage point decline ($p<.001$) in hit rates after the retraction event (Fig. 3B, right). Counting citations in the immediate two years after each publication also shows large negative impact effects comparing treatment and controls (Fig. S16).

Difference-in-differences analyses on a year-by-year basis reinforce these findings. Fig. 3C presents the year-by-year pivot rates, comparing treatment and controls, and shows a sharp increase in pivoting starting in the retraction year. Similarly, Fig. 3D shows a sustained decline in hit rates starting in the retraction year. Two-stage least squares regressions, with the retraction event as an instrument, further show that these "push" pivots predict substantial declines in impact (Table S6). Numerous robustness tests are considered in the SI using different citation measures and timing (Table S6, Fig. S16). We further consider a smaller case study of replication failures, rather than retractions, drawing on the landmark 2015 study of reproducibility in psychology [58], where 100 papers were quasi-randomly chosen for evaluation and 64 contained non-reproducible results. Deploying the same treatment and control method as for paper retractions, this smaller study provides confirmatory results for pivoting and impact (SI Section 2.6.1, Table S7). Altogether, we see pivoting increases and hit rate declines in response to these external shocks. These analyses further confirm the findings of the pivot penalty, now in response to external events that "push" treated authors into new areas.

Beyond push-type events, researchers may also be pulled into new areas when novel and important research questions emerge. This leads us to our second case study, analyzing how researchers shifted to engage the COVID-19 pandemic. The advent of the pandemic allows high-scale investigation of individual researcher pivots while further unveiling how science as a whole responds to a new and consequential demand upon the research community. Indeed, confronted by COVID-19, the world looked to science to understand, manage, and construct solutions, all in rapid fashion. Given that few researchers were studying coronaviruses or pandemics prior to 2020—and exactly zero were studying COVID-19 specifically—the emergence of COVID-19 called upon researchers across the frontiers of knowledge to consider a shift in their work [59-61] to address new and high-demand research questions.

Figure 4 investigates COVID-19 research and shows that pivoting to address COVID-19 was widespread. Although the earliest papers on COVID-19 did not appear until January 2020 [62, 63], by May 4.5% of all new scientific papers were related to COVID-19 (Fig. 4A). Further, while fields differed in their rate of pivoting, all fields pivoted to COVID-19 related research to some extent (Fig. 4B). Health sciences exhibit the greatest COVID-19 orientations, while social science fields – including economics, education, and law – also addressed COVID-19 relatively heavily, speaking to the pandemic's socioeconomic challenges [64, 65]. Furthermore, while fields inherently differ in their propensities to produce COVID-19 research (Figs. 4B, S17), we find that scientists in every field undertake unusually large pivots when writing COVID-19 related papers. Fig. 4C further tracks a cohort of scientists across the body of their work. It compares authors who wrote a COVID paper in 2020 and a control set of authors who did not, where control authors are matched to the COVID authors by cohort, field, and publication rate (see SI S2.7 for details). We find that pivot size presents a clear jump for COVID-related work, where COVID authors pivoted to an unusual degree compared to their own prior history, to their non-COVID 2020 papers, and to the control authors. In sum, unusually large individual pivots were a widespread phenomenon as scientists sought to address COVID-19.

We next turn to impact. Given that 2020 papers have had less chance to receive citations [66], we examine journal placement, where each journal is assigned the historical hit rate of its publications within its field and year (see SI S2.2). Fig. 4E considers all papers published in 2020, separating them into COVID and non-COVID papers. We find a large premium associated with COVID-19 papers, as reflected by the substantial upward shift in journal placement, consistent with the extreme interest in the pandemic. At the same time, the negative relationship between pivoting and impact persists and remains steep. Thus, scientists who traveled further from their prior work to write COVID-19 papers were not immune to the pivot penalty; rather they produced research with substantially less impact on average relative to low-pivot COVID papers. These results are also reflected in analyses net of individual fixed effects, where the pivot penalty is substantial though less strong than in the raw data (Fig. S18). Importantly, the pivot penalty is sufficiently steep that the COVID impact premium is mostly offset by the unusually large pivots associated with COVID research. For example, the upper 45% of COVID-19 papers by pivot size have lower average journal placement than non-COVID papers with median or smaller pivot size.

In sum, the "pull" nature of COVID-related work presents two extremely strong yet contrasting relationships regarding impact. On the one hand, this work has experienced an impact premium, consistent with the value of researching high-demand areas. On the other hand, greater pivot size markedly predicts less impactful work. These findings underscore a central tension for individual researchers and the adaptability of science in response to external opportunities: while working in a high-demand area has value, pivoting exhibits offsetting penalties.

**Pivots and Moderating Factors**

The pivot penalty presents a striking challenge in shifting research streams, both generally and in the context of external events, including in the pandemic where there is enormous societal demand and both scientists and science institutions worked hard to address the challenge. Indeed, in addressing the challenge, a key question is whether potential moderating factors might facilitate successful pivots. We close by studying possibilities informed by the science of science literature [67-69] and investigate further the sciences as a whole and the COVID-19 case study. First, what role may teamwork, including new coauthors, play in facilitating pivots? Second, what role does funding play? Finally, can these features overcome the pivot penalty?

Teamwork may be a critical feature in facilitating adaptability. Not only are teams increasingly responsible for producing high-impact and novel research [28, 33, 41, 70], they can also aggregate individual expertise [15], extending an individual's reach and promoting subject-matter flexibility [29, 71]. Indeed, looking to the pandemic, we find that team size was larger for COVID-19 papers than is typical in the respective field. Compared to field means, COVID-19 papers see 1.5 additional coauthors on average (a 28% increase in team size, Fig. S19). Further, COVID-19 authors work to an unusual degree with new coauthors (Fig. 4F), rather than existing collaborators, and engaging new coauthors is associated with larger pivots (Fig. S20). These results are consistent with teamwork expanding reach [15, 72, 73]. Nonetheless, we again see the pivot penalty in both large and small teams, and in teams with and without new coauthors (Fig. 4G-H). Thus, while bigger teams and teams with novel coauthors appear to predict higher impact, the pivot penalty persists.

We further probe adaptability through the lens of funding. We integrate funding data from Dimensions, which incorporates 600 funding organizations worldwide, and identify grant-supported research in 2020 for COVID and non-COVID papers (Fig. 4I). We see that grant-supported research disproportionately features small pivots. Specifically, there is a large and monotonic decrease in grant-supported research as pivot size increases, and this relationship is especially pronounced for COVID-19 papers, which are less likely to cite a funding source across all pivot sizes. These findings are natural to the extent that funding supports specific agendas, so that large pivots in general, and COVID-19 pivots in particular, tend to occur without acknowledging specific grants. Nonetheless, returning to impact, we find that the pivot penalty persists whether the paper does or does not acknowledge a specific grant, both in science as a whole and among COVID-19 research (Fig. 4J).

Altogether, one can consider numerous potential moderating factors and forms of heterogeneity, including individual's career stage and productivity, project-level team size, the use of new coauthors, and funding. When examining impact, however, we find that the pivot penalty persists regardless of these features. We further use regression methods to incorporate detailed controls for all these features together (see SI S2.5), finding that net of all these features, the pivot penalty moderates only slightly and remains substantial in magnitude (Fig. 4K).

**Discussion**

Science must regularly adapt to new opportunities and challenges. Yet the findings in this paper point to significant difficulties in adapting research streams, with implications for individual researchers, research organizations, and science and society as a whole. At an individual level, a researcher must weigh whether to continue exploiting a familiar research stream against opportunities that stand further away. Research on creativity suggests the value of exploration, novelty, and outsider advantages [22, 28-36], suggesting a risk vs reward tradeoff as researchers venture further from their prior expertise. However, other viewpoints emphasize the value of deep expertise, especially in drawing upon the frameworks, facts, and tools built by prior scholars [15, 45]. As Einstein observed, "…knowledge has become vastly more profound in every department of science. But the assimilative power of the human intellect is and remains strictly limited. Hence it was inevitable that the activity of the individual investigator should be confined to a smaller and smaller section" [48]. Consistent with Einstein's observation, as well as prior studies indicating increasing specialization and disadvantages when inventors switch fields [14, 15], we find that researchers face systematic challenges to pivoting research, and increasingly so with time. This 'pivot penalty' applies to both science and technology, generalizes across research subfields, and extends to the practical use and market value of ideas, external to the research domain. The pivot penalty further appears in response to external events that may push a researcher away from a given area or pull them into a new one. In the COVID-19 pandemic, the enormous demand for COVID-related research attracted numerous researchers and provided them an impact premium; yet we find that the pivot penalty continues to appear strongly among scholars who reached further to engage COVID research.

The pivot penalty, and its steepening with time, raises key questions for research organizations and research policy. For example, businesses and other organizations are often displaced by new entrants [50, 74], despite R&D efforts by the incumbents, which often fail to understand or embrace new technological opportunities [31, 37, 75]. The pivot penalty, uncovered in this paper, reinforces this organizational challenge and points towards tactics like "acquihires," where the research organization seeks to hire relevant experts rather than expect success by pivoting their existing personnel [76, 77].

More broadly, the pre-positioning of researchers appears to be a fundamental constraint on adaptability. In Louis Pasteur's famous words, "chance favors only the prepared mind," and without the pre-positioning of relevant human capital the coronavirus pandemic would likely have been still more costly. Portfolio theory points to diversified investments as a key tool to manage risk [78], but the pivot penalty suggests that, unlike typical investments, adjustments to the research portfolio are governed by substantial inertia [79]. From this perspective, investing explicitly in a diverse set of scientists becomes critical from a risk management standpoint. A diverse portfolio of investments can then play essential roles in both advancing human progress in ordinary times [22, 80] while also expanding human capacity to confront novel challenges.

Science and technology present evolving demands from many areas – from artificial intelligence to genetic engineering to climate change – creating complex issues, risks, and urgency. This paper shows that pivoting research is difficult, with researchers' pivots facing a growing impact penalty. It is notable that the pivot penalty not only appears generally across scientific fields and patenting domains, but also appears amidst major events in science, including when prior areas become devalued, as in the retraction context, and when high-demand areas emerge, as in the COVID-19 pandemic. Nevertheless, studying adaptability in different settings and time scales, including longer-run research shifts, are key areas for future work. For example, should a researcher give up in the likely event of a failed pivot or alternatively further develop their expertise in the new area and stick to the new path? Understanding such sequential dynamics may not only reveal further insight about the nature of creative search [34, 35] but also help us better understand how to create conditions to facilitate adaptive success. Lastly, pivoting to address new challenges is not unique to science and technology but may underpin the dynamics of success and survival for individuals, firms, regions, and governments across human society [9, 75, 81-84], suggesting the uncovered pivot penalty may be a generic property of many social and economic systems, with potential applicability in broader domains.

**Acknowledgements** We thank seminar participants at University College London, the University of California Berkeley, Brigham Young University, Georgetown University, Hunan University, Northwestern University, Yale University, and conference participants at the American Economic Association annual meetings for invaluable comments. We gratefully acknowledge grant support from the Air Force Office of Scientific Research under award number FA9550-17-1-0089 and FA9550-19-1-0354, the National Science Foundation under award number SBE 1829344, the Alfred P. Sloan Foundation under award number G-2019-12485, and the Peter G. Peterson Foundation under award number 21048.

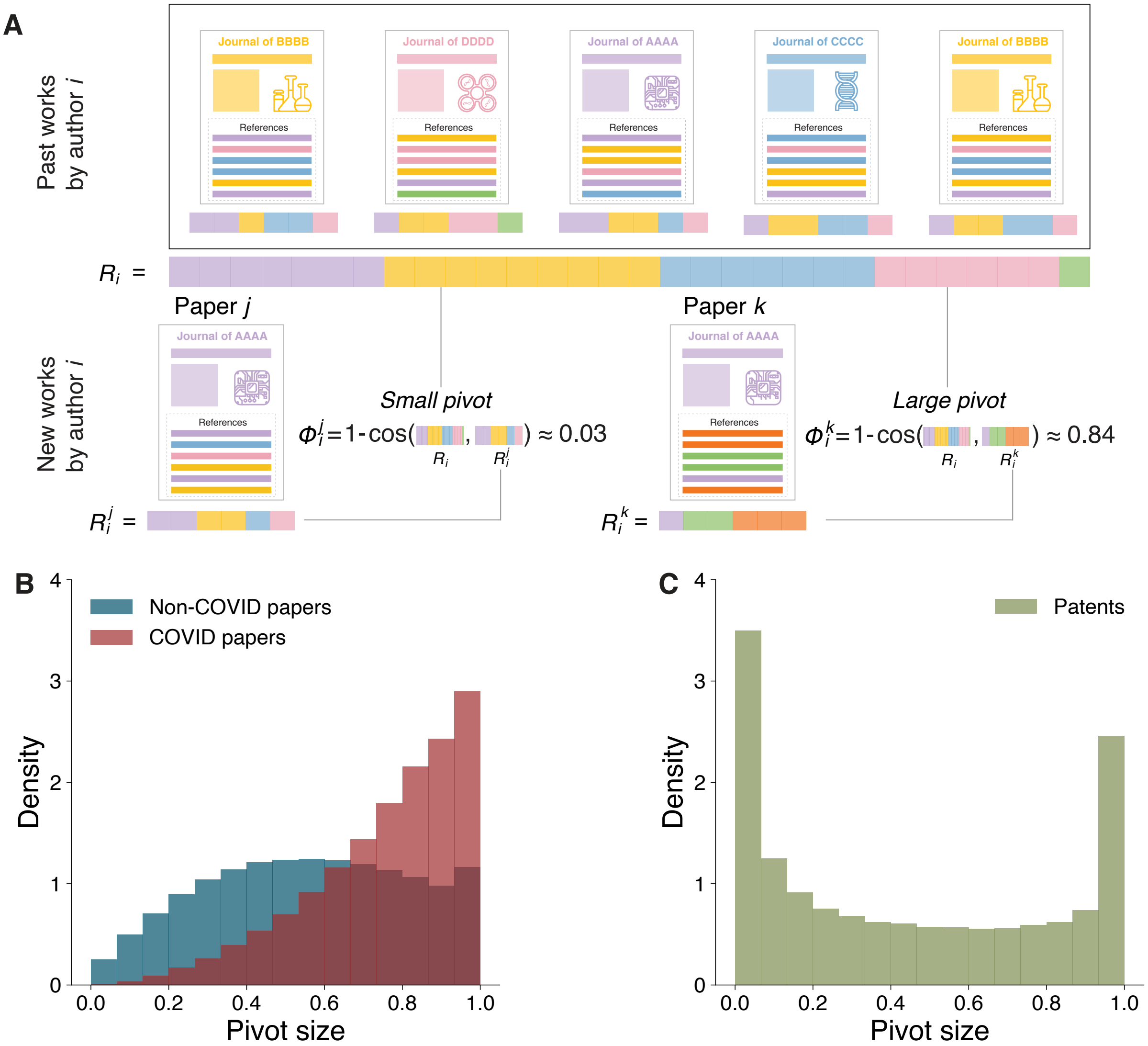

**Figure 1. Quantifying Research Pivots.** (**A**) The pivot measure compares a focal work against prior works by the same researcher. An increasing value on the [0,1] interval indicates a larger pivot from the researcher's prior work. In the sciences, journals are used to define research areas (pictured); in patenting, technology classes are used. The distributions of pivots in 2020 show wide dispersion in science (**B**) and in patenting (**C**). COVID-19 papers show higher pivots (**B**) than other papers in 2020.

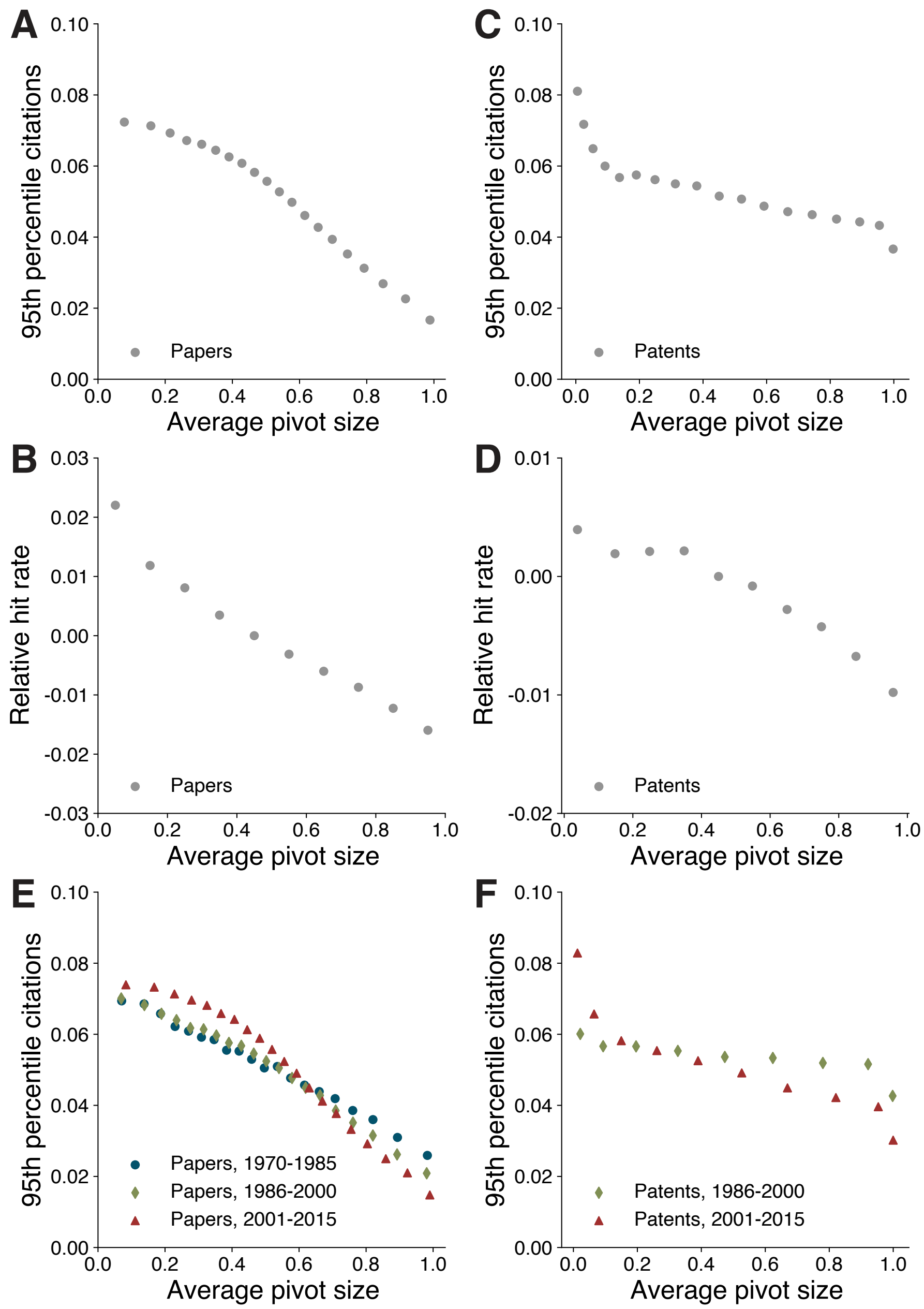

**Figure 2. The Pivot Penalty. (A)** Studying 26 million papers published from 1970-2015, papers with higher pivot size present substantially lower probabilities of being high impact. (**B**) Further, for a given author, relative impact among their papers declines steeply with pivot size. (**C**) Studying 1.8 million U.S. patents granted from 1980-2015, patents with higher pivot size present substantially lower probabilities of being high impact. (**D**) Further, for a given inventor, relative impact among their patents declines with pivot size. Over time, the relationship between pivot size and high impact works has become increasingly negative in science (**E**) and patenting (**F**).

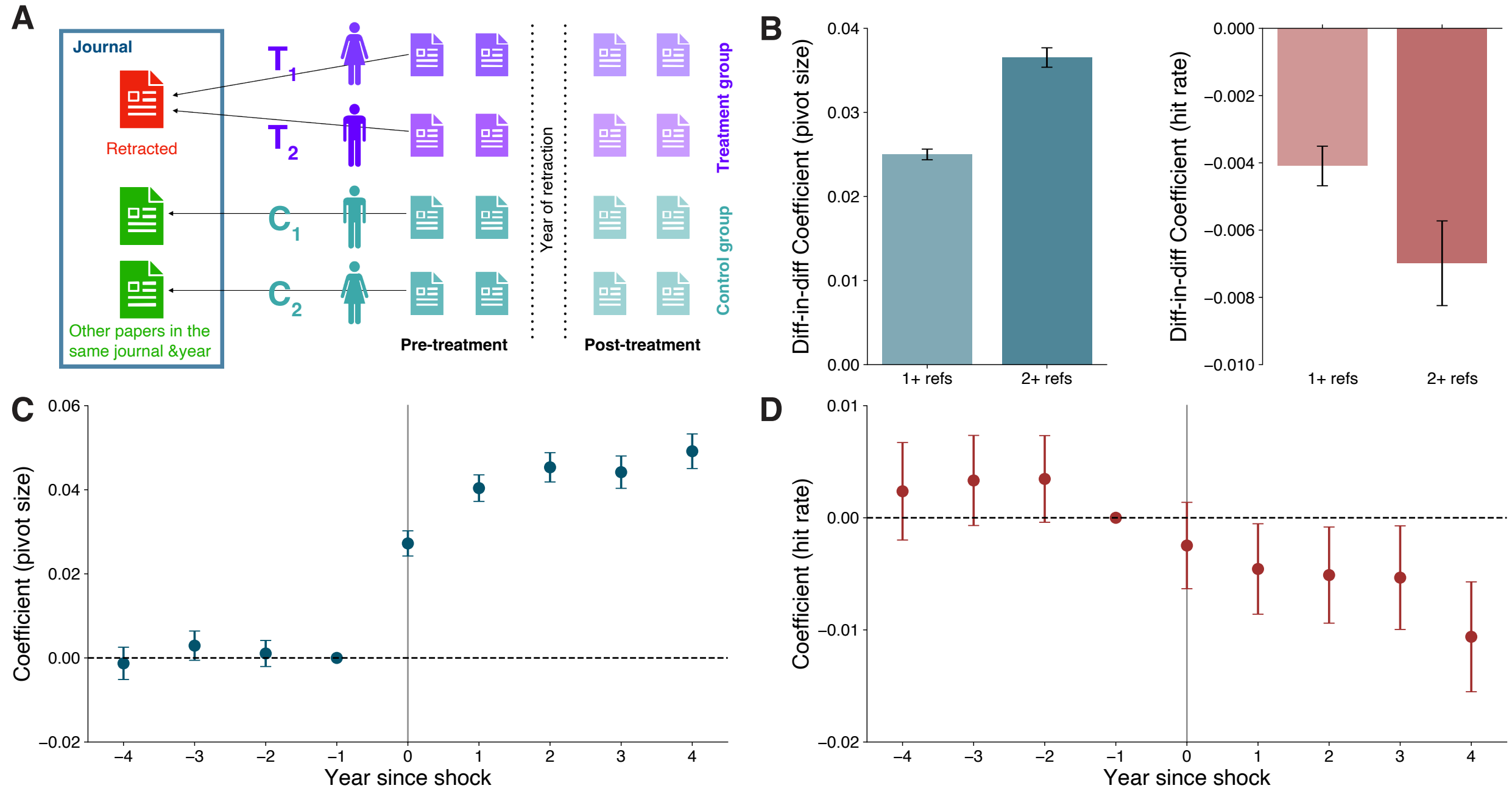

**Figure 3: Pivots and Retraction Events.** (**A**) The difference-in-differences analysis compares treated scientists who directly cite a paper prior to its retraction to control scientists who cited other papers in the same journal and year as the retracted paper. Pivot size and impact of treated scientists is compared to control scientists before and after the year of retraction. (**B**) Pivot size significantly increases for treated scientists relative to control scientists after the retraction. The effect is larger when focusing on scientists who cited the retracted paper at least twice. Hit rates fall for treated scientists after retraction, and again the effect is stronger for those citing the retracted paper at least twice. A year-by-year analysis again shows a marked increase in pivot size (**C**) and a decrease in hit rate (**D**) for treated authors relative to control authors, starting in the retraction year.

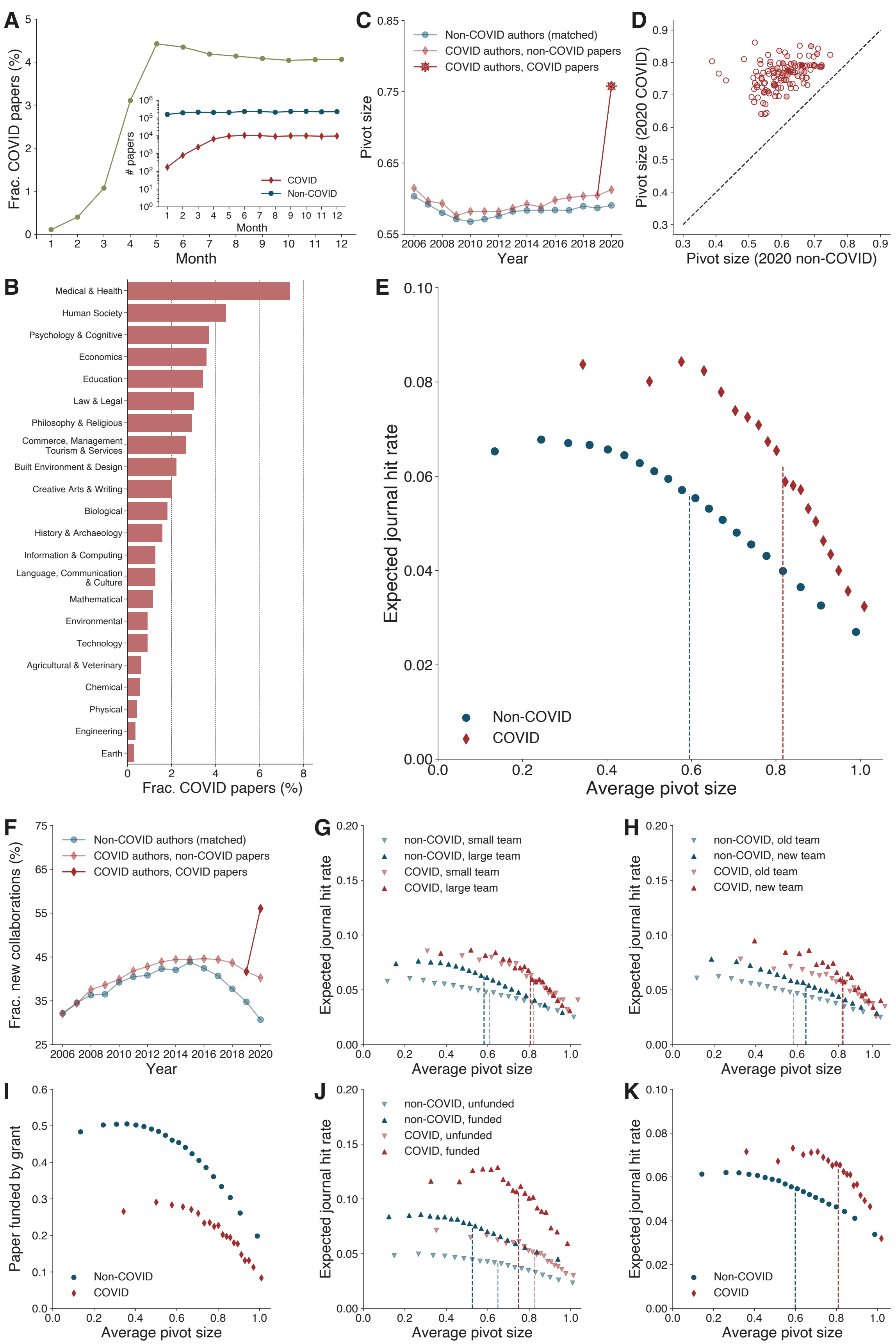

**Figure 4. Pivots and the COVID-19 Pandemic.** (**A**) Science rapidly shifted to COVID-19 research in 2020, with COVID-19 publications rising to 4.5% of all science publications in May 2020 and maintaining similarly high rates thereafter. (**B**) While health sciences and social sciences featured the strongest responses, all scientific fields engaged COVID-19 research. (**C**) Comparing COVID and non-COVID papers within each field in 2020, unusually large pivots have been a universal feature of COVID-19 research in all 154 subfields of science. (**D**) Scientists who write COVID-19 papers pivot to a greater extent than they do in their prior work, their other 2020 work, or matched control scientists' do. (**E**) COVID-19 papers experience an impact premium, but the pivot penalty appears within both COVID and non-COVID work. Comparing at the median pivot sizes (dashed lines), the COVID-19 impact premium is substantially offset by the pivot penalty, given its larger median pivot size. (**F**) Engaging new collaborators was especially common for COVID-19 researchers, who worked with new collaborators to an unusual degree compared to their own prior history, their other 2020 publications, and control scientists. Nonetheless, the pivot penalty persists for big and small teams (**G**) and when engaging new or existing coauthors (**H**). (**I**) Funding support is heavily oriented to lower pivot work. Higher-pivot work is substantially less likely to acknowledge funding support in the sciences as a whole (blue) and among COVID-19 papers (red). COVID-19 papers were especially unlikely to acknowledge grant support. Yet the pivot penalty appears even among both funded and non-funded work (**J**). (**K**) While individual, collaborative, and funding features sharply condition the adaptive response of science, in regression analysis they do not individually or collectively overcome the fundamental pivot penalty.

# Supplementary Information for
# Adaptability and the Pivot Penalty in Science and Technology

Ryan Hill[1,2¶], Yian Yin[1,3,4,5¶], Carolyn Stein[6,7], Xizhao Wang[2], Dashun Wang[1,2,3,4*], Benjamin F. Jones[1,2,3,8*]

[1]Center for Science of Science and Innovation, Northwestern University, Evanston IL

[2]Kellogg School of Management, Northwestern University, Evanston IL

[3]Northwestern Institute on Complex Systems, Northwestern University, Evanston IL

[4]McCormick School of Engineering, Northwestern University, Evanston IL

[5]Department of Information Science, Cornell University, Ithaca NY

[6]Haas School of Business, University of California – Berkeley, Berkeley CA

[7]Department of Economics, University of California – Berkeley, Berkeley CA

[8]National Bureau of Economic Research, Cambridge MA

[¶]These authors contributed equally to this work.

[*]Correspondence to: dashun.wang@northwestern.edu, bjones@kellogg.northwestern.edu.

## S1 Data description

### S1.1 Publications data

Our primary dataset for scientific publications is based on Dimensions, a data product from Digital Science[1,2]. Dimensions is one of the world's largest citation databases, including scientific publications from journals, conference proceedings, books and chapters, and preprint servers. For each paper we obtain its title, publishing venue, list of authors, affiliation(s), publication date, fields of study, references, number of citations received, and acknowledged funding information. We collect all publications in the Dimensions database through December 31, 2020, and our analysis covers publications over the 1970-2020 period.

Our baseline dataset is restricted to the 45.2 million papers that include at least five references to prior work. We do this for two reasons. First, pivot size — our primary variable of interest — uses reference information as a proxy of a paper's knowledge sources. Second, this restriction helps filter out non-research articles and incomplete records. A manual check of a random sample of excluded papers (those with 4 or less references to prior work) shows that many are commentary or editorial pieces that cite very few references, while some are due to the lack of reference data sharing between Dimensions and some publishers. As robustness checks, we also consider further analyses where reference counts exceed successively higher thresholds, including at least 15, 20, 30 or 50 references to prior work (see SI Section S3).

Recent studies have shown that Dimensions covers most reference-citation linkages as recorded in other bibliographic databases, such as the Web of Science or Scopus[3]. For publications with preprint linkages, we further combine the published and preprint article into a single record and count citations as the sum of citations to the combined record. For most journal articles (99.2%) and all preprint publications, we have information on its publishing venue (i.e., specific journal or preprint series).

<u>Author Name Disambiguation</u> An important step in science of science studies is author name disambiguation. Dimensions has developed a systematic algorithm using both internal information derived from papers (e.g., affiliation and citations) as well as external author profile information (e.g., ORCID). Within our sample, 84.6% of records in the data are assigned author IDs. Analyses are restricted to papers with at least one disambiguated author. We further hand check random samples of papers and author IDs against public CVs, which indicates high quality matching of authors and their papers (see SI Section S3).

<u>Affiliation disambiguation</u> Dimensions has also mapped raw affiliation strings to the Global Research Identifier Database (GRID), which provides unique identifiers for different research organizations, offering unique affiliation IDs to 75.1% of records in the data.

<u>Field classification</u> Dimensions also implements classification approaches to assign fields under the FOR (field of research), a two-level field classification system (22 level-0 fields and 154 level-1 fields). Across all of Dimensions, 94.3% and 88.6% of papers have at least one level-0 and level-1 field, respectively. The median number of level-0 (L0) fields per paper is 1 and the median number of level-1 (L1) fields per paper is 1. In analyses where we calculate field-specific means, we associate papers to each of the fields they belong to, when the paper has multiple fields.

Retracted publications For the retraction natural experiment, we start with the complete list of retracted papers produced by the Retraction Watch Database and CrossRef. This retractions data has become open source and available through CrossRef[4]. We merge retracted papers based on DOI to the Dimensions database (through 2020). This merge identifies 13,456 retracted papers in the Dimensions database over the 1975-2019 period. See Methods below for the construction of the treatment and control groups and further details.

Replication failures As a substantially smaller case study, we use the landmark replication analysis in 2015, “Estimating the reproducibility of psychological science” [5]. Specifically, we take all publications in Dimensions that appear in three journals (*Psychological Science*, *Journal of Personality and Social Psychology*, and *Journal of Experimental Psychology: Learning, Memory, and Cognition*) in 2008, which were the source of the 100 publications that were quasi-randomly selected and analyzed in the replication study. Of the tested papers, 64 had results that failed to replicate. See Methods below for the construction of the treatment and control groups and further details.

COVID-19 related publications We constructed a set of COVID-19 related publications using a keyword search method and following previous work[6] by searching for papers published in 2020 with the following query:

*"2019-nCoV" OR "COVID-19" OR “SARS-CoV-2” OR "HCoV-2019" OR "hcov" OR "NCOVID-19" OR "severe acute respiratory syndrome coronavirus 2" OR "severe acute respiratory syndrome corona virus 2" OR (("coronavirus" OR "corona virus") AND (Wuhan OR China OR novel))*

Since our primary interest is papers closely related to the COVID-19 pandemic, we limit the search to the title and abstract, yielding 95.5 thousand COVID-related papers.

### S1.2 USPTO patents data

We further leverage USPTO patents to test the pivot penalty in technological areas. Our data comes from bulk data services provided by PatentsView (retrieved in March 2021), a patent data platform supported by USPTO. The original data covers 6.9 million patents granted by USPTO since 1976, with detailed information on the patent's title, date, CPC classification codes, patent references, and (disambiguated) inventors.

For our analysis, we focus on utility patents. Given that we are primarily interested in career-level pivot behaviors, we exclude all patent continuations (by definition, they will be highly similar to previous patents by the same inventor). To this end, we further retrieve the application numbers of all patents in our dataset and link them with continuation information from the Patent Examination Research Dataset (PatEx). We then remove all patents that are associated with at least one "parent" patent in continuation records. Consistent with our selection on publication data (S1.1), we further focus on patents with at least 5 patent references. This results in a subset of 3.7 million patents, and 3.0 million patents of over our main analysis period, 1980-2015.

Inventor Name Disambiguation PatentsView has also developed an inventor name disambiguation algorithm, which assigns a unique id for each inventor record in the database. More details about the algorithm are available at https://github.com/PatentsView/PatentsView-Disambiguation. We use this information to construct inventor career trajectories. See Methods below.

Technology classification We use the Cooperative Patent Classification (CPC) to determine the technology class of each patent. The CPC system applies a five-level hierarchical system: (a) 9 sections (e.g., B), (b) 128 classes (e.g., B29), (c) 662 subclasses (e.g., B29C), (d) 9,987 groups (e.g., B29C45), and (e) 210,347 subgroups (e.g., B29C45/64). The subset analyzed in Fig. 1 and 2 are associated with 6.0 classification codes per patent on average. For calculation of cosine similarity and pivot size, we pool together the classification codes associated with each reference.

### S1.3 Scientific grant data

We also use scientific grant data from Dimensions. Dimensions collects over 5 million granted projects from over 600 funders across the world. For each project, the data includes the project title, investigators, funder, funding amount, internal project number from the funder, and project duration. Name disambiguation for the investigators and publication authors shares the same ID system in Dimensions, allowing us to examine the funding situation of each author. Here we focus on all grants with end dates no earlier than 2019 to approximate the set of recently funded investigators/authors. In addition, Dimensions combines funding and publication records as well as text mining from acknowledgement statements to infer whether a paper is supported by a funder or a specific grant.

### S1.4 Patent References to Science

Going beyond paper-to-paper citations, we measure the patent impact of scientific papers using Reliance on Science data[10] (v34). The original linkage contains 40.4 million citation pairs from worldwide patent documents to scientific papers indexed in the Microsoft Academic Graph (MAG). In our analysis, we focused on USPTO patents. We further merge the cited papers to the Dimensions database using DOI. Together, these steps yield 4.1 million Dimensions papers with at least one USPTO patent citation.

### S1.5 Market Value of Patents

Data for the market value of patents is from Kogan et al. (2017)[7]. Their study considers stock price reactions on the day patents are issued to the relevant firm. Using an event study design, they produce a market valuation for each patent. This method, and hence the data, only applies to patents from publicly traded firms. Market valuations and pivot size are available for 802 thousand patents issued over the 1980-2015 period.

## S2 Methods

### S2.1 Individual careers

Our analyses focus on the pivoting behavior of researchers and resulting impact. To measure pivoting behavior, we need a stream of work by that researcher so that a given work can be compared to the prior work of that researcher. We therefore focus on researchers who have at least 5 works. Any papers for which a researcher is an author, including coauthored papers, are considered part of that person's stream of work. Similarly, coinvented patents are attached to each coinventor. For papers, the resulting dataset includes 37 million papers over the 1970-2020 period. For patents, where inventors tend to be less prolific and having at least 5 patents is not as common, the resulting dataset includes 1.8 million patents over the 1980-2015 period. For both papers and patents, we use the name disambiguated identifiers provided by our data sets (see above).

Given the body of work assigned to each researcher, we further calculate several relevant metrics. These include career age, which is measured as the number of years since the author's first publication year (or, for inventors, the number of years since the first patent application year). We further calculate the citation impact of each researcher's individual works, their modal field of research, their number of publications, and their typical pivot size. These characteristics can be further used to define control groups when studying how researchers respond to external events.

### S2.2 Pivot Size

We calculate pivot size as described in detail in the main text. Figure 1 presents the pivot size distributions for papers (Fig. 1B) and patents (Fig. 1C). Here we consider additional assessments of pivoting behavior and pivot size distributions, using alternative classifications in the data, and further demonstrate the robustness of the pivot penalty to alternative approaches.

#### S2.2.1 Papers

For papers, our main analyses use the prior three years of publications by a given author to calculate pivot sizes for any given paper. One can alternatively use all the prior works of the author, at the given point in the career, to calculate pivot sizes. Fig. S1 presents the pivot size distribution using the full list of prior work and shows that pivot size distribution remains similar. Further, the pivot penalty continues to appear (Fig. S1). As an additional robustness test, these further analyses also examine COVID-19 research and continue to find similar findings as when using the three-year window.

An alternative and substantially coarser way to calculate pivots for papers is to use fields instead of journals. Specifically, one can code a paper's references not by their journal but by their L1 field code and then rebuild the pivot measure for each paper on this basis. This is a much coarser approach, in the sense that there are 154 L1 fields, while there are 40,225 journals. In terms of the number of fields, coding papers in this coarser way is akin to the level-two CPC coding variant for patents (see below).

Fig. S3 compares pivot size as measured using L1 field codes with pivot size when measured using journals. We see a monotonic positive relationship. Note also that, using the field codes, the pivot measure is compressed to lower values, with a pivot size above 0.5 accounting for only 5 percent of papers. This leftward shift in the pivot size distribution is natural when using a coarser knowledge coding, as shifts in referenced fields are bigger and rarer. Fig. S3 further shows that we continue to see a pivot penalty when using L1-field codes to define pivot sizes, with the pivot measure compressed to lower pivot sizes. The main text features the journal-based analysis, both because it shows greater range in pivoting and because of known concerns about the quality of field encodings in paper data (Bornmann 2018), which may lead to noise and attenuation when relying on the field encodings.

S2.2.2 Patents

For patents, one can draw on the hierarchical nature of CPC technology classification to re-examine pivot sizes and the robustness of the pivot penalty. Fig. S2 shows that, as one shifts from the narrowest technology classification (level-5, with 210,347 subgroups) to the broadest (level-1, with 9 sections), the pivot distribution for inventors shifts leftward. As with papers, coarser field classifications naturally result in leftward shifts in the pivot size distribution. Nonetheless, pivoting behavior remains highly variant. Fig. S4 further shows that, regardless of the technological classification used, the pivot penalty is robust in all cases. Thus, for the papers and patents, the fundamental finding of a pivot penalty endures regardless of numerous alternative coding schemes for areas of knowledge.

A distinction for patents, compared to papers, is that the pivot size distribution appears bimodal, with a tendency towards very small pivots and very large pivots, and indeed this feature endures using broad or narrow field encodings (Fig. S2). The high frequency of large pivots (near 1) is due in part to cases where there are relatively few references in prior art from a given inventor. This could occur either because the inventor has few prior patents or because that inventor's prior patents make few prior art references. To further investigate this dimension, Figure S23 presents the patent pivot size distribution when we restrict the sample to inventors with at least 10 patents in the prior three years. Figure S23 further presents the patent pivot size distribution when we restrict the sample to inventors with exactly one patent in the prior three years, but then separate out cases where that patent has at least 100 prior art references. We see that in both cases the presence of very high pivot patents declines substantially. That said, the bimodal nature of the patent relationship still remains and appears robust to these reference count considerations. A substantive reason for the bimodal behavior of inventors may be that inventors in patenting contexts are often assigned to their research directions. For example, based on the corporate priorities, inventor R&D groups may be asked to engage new areas (high pivots) or double down on existing areas (low pivots) following the direction of R&D management and the interests of the firm.

### S2.2.3 Manual Checks for High Pivots

We studied high pivots to see if very large pivots may be related to any name disambiguation issues. Specifically, we took a random sample of 10 authors who produce a paper with a pivot score >0.95 in the year 2020. We then took this very high pivot paper (10 papers) as well as all other papers in the database that were associated with that author and published over the prior three years (totaling another 148 papers). We then hand checked every paper associated with these authors against the authors' CVs, personal websites, Google Scholar profile, PubMed page, or Scopus page (depending on the source available for a given author). The large majority of the 158 papers, including all 10 very high pivot papers, could be verified as matches through the authors' own CVs/websites/Google scholar etc. profiles. The very high success rate gives further confidence that name disambiguation is sufficiently accurate. See Section S3.1 below for further analysis and detail.

### S2.2 Impact measures

S2.2.1 Main Citation Based Measure

Our primary measure of paper impact is an indicator for whether an article is in the 95th percentile or higher of citations compared to articles published in the same year with the same L1 fields. Similarly, we define a measure of patent impact by looking at whether a patent is in the 95th percentile or higher of citations compared to patents in the same year with the same primary class. This approach provides a binary outcome variable, where 1 indicates a high impact work and where the mean hit rate is 5% by construction in any given field and year in the data. As such, this method normalizes the outcome across different fields and across different periods of observation.

S2.2.2 Alternative Citation Based Measures

Our main analyses focus on a binary indicator for being especially high impact, with two primary motivations. First, scientific and technical progress may hinge on key, high impact ideas, and second, theories of creative search (such as explore vs. exploit frameworks, or emphases on creative outsiders) often orient on the production of high-impact ideas. When looking within individual careers, however, high impact work can be rare, and many individual researchers do not produce high impact works. Thus, smoother outcome measures of impact may be useful especially when looking within individual careers, as well as for providing robustness checks on broader findings. We therefore consider numerous additional citation measures, which can be used to further characterize results.

*Mean citations*. As a smoother measure of the citation impact, and following prior literature, we take a paper's citation count and normalize it, dividing by the mean citation counts for papers in that field and with the same publication year[8,9]. This approach allows for a more continuous measure of impact while also continuing to normalize for citation differences across fields and time. Figure S9 shows that the pivot penalty remains large using this more continuous measure. Specifically, the citation impact of the lowest pivot papers tends to be approximately 30% above the field-year mean while the citation impact of the highest pivot papers tends to be approximately 55% below the field-year mean. Similarly, looking within individual careers, Figure S9 shows substantial pivot penalties net of individual fixed effects.

*Mean percentile citations*. Another approach to measuring impact converts each paper to its percentile of citations received among all papers published in that field and year. This provides another, smoother version of citation impact compared to the binary measure, while also limiting the influence of any outliers. Figure S9 shows that the pivot penalty continues to appear and remain substantial using this alternative measure. Specifically, the lowest pivot papers on average are in the 58$^{th}$ percentile of citations received compared to other papers in the field and year, while the highest pivot papers on average are in the 36$^{th}$ percentile.

*Time frame*. Comparing works published in the same field and year acts to normalize the measure of citation impact. That is, citation impact is always being compared among papers in the same field and with the same horizon for citation (between the publication year and the present). However, one can also use a fixed period after publication as an additional way to normalize the time frame. We recompute citations received by each paper using, alternatively, two-year, five-year, and ten-year forward windows. Figure S8 shows that the pivot penalty is similar regardless of these alternative citation windows.

*Alternative binary indicators*. Within the class of binary indicators, which are useful to emphasize the locus of the highest-impact work, one can consider alternative percentile thresholds. The main analyses consider papers with citations received in the 95$^{th}$ percentile or above. Alternatively, we consider indicators for work in the upper 90$^{th}$, 95$^{th}$, 99$^{th}$, 99.5$^{th}$ and 99.9$^{th}$ percentile of citations received. Figure S13 shows that the pivot penalty remains severe at these different impact thresholds, and indeed high-pivot work is even less likely to achieve the higher impact thresholds compared to low pivot work, as discussed in the main text.

S2.2.2 Non-Citation Based Measures

*Journal Placement* When examining research in 2020 (including COVID-19 research) there has been less opportunity for works to accumulate citations. We therefore use journal placement as an alternative. For the journal impact measure, we mirror the baseline citation measure by calculating the share of papers in a given journal that reach the 95$^{th}$ percentile of citations (within its field and year), averaged between 2000-2019. This metric is intended to infer both the likelihood of becoming a hit paper and the authors' perception of a paper's impact and contribution based on journal placement.

*Publication Success* Our datasets consider published papers and granted patents, following standard practice in analyzing science and innovation outcomes. However, for recent years in science, we can also take an additional step by using preprints and asking whether preprints ever become published. Specifically, we examine all 1.07 million preprints released from 2015-2018 on preprint databases such as arXiv and SSRN. We define an indicator variable that is equal to 1 if the working paper becomes a published article in our data.

Fig. S11 shows that higher pivot sizes are associated with a large decline in the probability of being published. There is a smooth decline in publication propensity with pivot size, where virtually all of the lowest-pivot size working papers become published while only two-thirds of the highest-pivot papers are published (within five years). Thus, not only do higher pivots exhibit declining impact, conditional on publishing, high pivots are also harder to publish. The impact challenges of high pivot appear only stronger from this additional viewpoint. For example, taking $\Pr[\text{Hit}] = \Pr\,[\text{Hit}|\text{Published}] \times \Pr[\text{Published}]$ (where we note that unpublished papers cannot be hits), the highest-pivot papers would see unconditional hit rates that are approximately two-thirds lower than the (already low) hit rate conditional on successfully publishing. This method of observing preprints that fail to publish opens new avenues of insight in the science of science that may be used in other studies.

*Patent Usage of Science* As an outcome measure for a given scientific work, we further consider whether the article is referenced as prior art in a future patent. This outcome indicates applied use of the scientific ideas, suggesting an important form of impact that occurs beyond the domain of science itself [10,11]. For each scientific article, we use a binary indicator where 1 indicates that the article is directly referenced in a future patent. Fig. S10 shows that high-pivot papers are substantially less likely to be referenced in a future patent. The pivot penalty is not monotonic in this case, with low pivot sizes associated with less applied use. However, this outcome measure is not normalized by scientific field (unlike the citation impact measure in science), and there is substantial heterogeneity across scientific fields in the frequency of direct patent citations[11]. For example, nanotechnology papers are far more likely to be cited in patents than astrophysics papers are. To absorb field heterogeneity, we further consider patent citations to science in a regression with fixed effects for the paper's L1 research field. The result, shown in Fig. S10, indicates that the non-monotonicity at low pivot sizes largely disappears with this simple field control.

*Market Value* As an outcome measure for a given patent, we consider the market value measure[7], which assigns to a patent a market value based on the stock price response of the business on the day of the patent's issuance. This measure is only available for patents that are assigned to publicly held firms. Fig. S12 shows that higher-pivot patents have substantially lower market value.

### S2.3 Tail Novelty and Median Conventionality

Following the existing literature, we measure the novelty and conventionality of academic papers by considering the combinations of existing ideas.[12] We consider the pairwise combinations of journals cited by each paper and compare them to the expected frequency that those combinations would appear by chance according to the existing network of citations. Two journals that are unlikely to be paired by chance indicate a novel combination, while journals that are more likely to be combined indicate a conventional combination. Specifically, a journal combination can be assigned a z-score, comparing the observed and expected frequency of that journal pairing in science and normalizing by its standard deviation. Each paper has a distribution of z-scores across all the journal pairs it references. Following prior literature, we denote a "high tail novelty" paper if its $10^{th}$ percentile z-score is below 0, and we denote a "high median conventionality" paper if its $50^{th}$ percentile z-score is in the upper half of all papers.[12] Past literature suggests that high impact papers typically have both high tail novelty and high median conventionality. Much of the work that those papers build on is rooted in conventional knowledge combinations but sprinkles in new combinations of existing knowledge as well[12]. In this paper, we use a pre-calculated version of novelty and conventionality provided by SciSciNet, an open source database[13].

Figure S14 shows that higher pivot papers tend to make novel combinations. This indicates that when researchers pivot, they are not only being novel in their own terms, compared to their own portfolio of research, but they also tend to introduce combinations that are novel in science as a whole. By contrast, high pivots are associated with low conventionality. Thus high-pivot work tends not to feature the grounding in conventional combinations that are a key combinatorial ingredient in predicting high impact work. Table S9 further considers regression analysis of pivoting behavior, novel combinations, and conventional combinations together. These regressions continue to show a substantial pivot penalty. Thus, while high pivots are strongly associated with low conventionality, regressions put the weight of impact on pivoting.

### S2.4 New collaborators

To track a scientist's engagement with new collaborators over time, we first construct a set of collaborators for each author-paper pair, tracking coauthorship interactions among all the disambiguated authors in the author set (see S2.1). We then sort all publications in one's career by publication date and sequentially calculate the number of new collaborators in each paper.

To further understand the characteristics of these new coauthors, we calculate their major field of research (both level-0 and level-1) before the paper's publication year. We also compare the affiliation information of new collaboration pairs (based on disambiguated GRID ID) to see if the focal author and collaborator share at least one common affiliation. Together, these measurements allow us to count the number of new collaborators, as well as whether new collaborators come from the same or different field or affiliation. If either the focal author or the collaborator has missing data in field or affiliation, this pair is considered as "unknown" and excluded in the same/different categorization.

### S2.5 Regression methods

<u>S2.5.1 Output Level Analyses</u>

Our most basic analyses consider the output level, where an observation is a given work (paper or patent) and where pivot size used is the mean pivot size among the members of the team. These regression models in general take the form:

$$Impact_i = \alpha + f(Pivot_size_i) + \boldsymbol{\theta X_i} + \varepsilon_i$$

where i indexes a given work (paper or patent), $Impact_i$ is one of the various outcome measures (see above), and $\boldsymbol{X_i}$ is a vector of control variables. Rather than imposing a linear or other functional form on the data, we write $f(Pivot_size_i)$ to emphasize potentially general functional forms for the relationship between pivot size and impact.

To reveal potentially non-linear relationships between pivot size and outcome variables, many analyses use binned scatterplots[14]. In Fig. 2A for example, we order the sample of papers by average pivot size along the x-axis and split the observations into 20 evenly-sized groups. Then each marker is placed at the mean (x,y) value within each group. Similarly, in Fig. 2C, we consider the same using patents.

We also extend the binned scatterplots analysis to include control variables. For example, for the multivariate regression results presented in Fig. 4K, we consider numerous additional controls, including fixed effects for average prior impact groups, author age groups, team size, the number of new collaborators, and an indicator variable for whether the paper was funded. To include regression controls while maintaining the non-parametric advantages of the binned scatterplot approach, we in practice run two regressions to residualize pivot size and impact, net of the controls, following the Frisch-Waugh-Lovell theorem. Specifically, we first run regressions of the form:

$$\text{PivotSize}_i = \alpha_1 + \boldsymbol{\theta_1 X_i} + \varepsilon_{1i}$$

$$\text{Impact}_i = \alpha_2 + \boldsymbol{\theta_2 X_i} + \varepsilon_{2i}$$

And then consider the binned scatterplot relationship between residual impact ($\widetilde{\text{Impact}}_i = \text{Impact}_i - \hat{\alpha}_2 - \boldsymbol{\hat{\theta}_2 X_i}$) and residual pivot size ($\widetilde{\text{PivotSize}}_i = \text{PivotSize}_i - \hat{\alpha}_1 - \boldsymbol{\hat{\theta}_1 X_i}$).

Finally, when looking at subsets of the data to test the robustness of a negative slope between pivot size and impact, we also consider the linear version of the baseline regression, taking $f(Pivot_size_i) = \beta Pivot_size_i$. For example, we run this regression separately in each L1 field for papers, and in each CPC patent class to test how often the negative slope of the pivot penalty appears. See Tables S1 and S2.

S2.5.2 Researcher Panel Level Analyses

To examine the relationship between pivot size and impact within individual researchers, we use a panel data structure. Observations are at the researcher-by-paper and researcher-by-patent level, which allows the inclusion of individual fixed effects. By including individual fixed effects, the regressions compare variation in impact within the individual against variation in pivot size within the individual. More generally, the individual fixed effects account for any fixed characteristic (observed or unobserved) for a given researcher. We will also use this panel structure, with individual fixed effects, when considering the natural experiment described further below.

The panel regression with individual fixed effects in general takes the form:

$$Impact_{ipt} = \mu_i + \gamma_t + \beta f(Pivot_Size_{ipt}) + \boldsymbol{\theta X_{ipt}} + \varepsilon_{ipt}$$

where *i* indicates a given researcher, *p* indicates a given work (paper or patent), and t indexes the year (publication year for a paper and application year for a patent). The $\mu_i$ are individual fixed effects, the $\gamma_t$ are time fixed effects, and $\boldsymbol{X_{ipt}}$ is a vector of other potential control variables. As before, we allow for potentially non-linear relationships between pivot size and impact and hence take a non-parametric approach. Specifically, we generate bins of pivot size and include indicator dummies for a work appearing in the relevant bin. Given the very large number of individual fixed effects, we run these models in Stata using reghdfe command suite[15]. Standard errors are clustered at the researcher level.

To analyze pivoting in more specific contexts, we take subsets of the data. For example, to shed light on reputational mechanisms, we consider the subset of data where an author has multiple papers in the same year in the same L1 field, or in the same year and in the same journal. Table S5 presents these results. To consider the effects of external shocks we take the subset of treated and control researchers (see next section).

### S2.6 Differences-in-Differences Methods

When studying external shocks, we continued to use the researcher panel data model with individual fixed effects. We implement standard difference-in-difference methods, comparing treated researchers to control researchers, before and after the external event. The regressions take the form:

$$Pivot_Size_{ipt} = \mu_i + \gamma_t + \beta Treat_Post_{ipt} + \gamma Post_{ipt} + \varepsilon_{ipt}$$

$$Impact_{ipt} = \mu_i + \gamma_t + \beta Treat_Post_{ipt} + \gamma Post_{ipt} + \varepsilon_{ipt}$$

where $Post_{ipt}$ is an indicator for the period after the shock. The indicator for being in the treatment group is absorbed with an individual's fixed effect and so does not appear separately in the regression. $Treat_Post_{ipt}$ is an indicator for being in the treatment group after the shock and provides the reported difference-in-differences estimate. The implications of the external event for pivot size and the reduced form results for impact are both shown in Fig. 3B. We also show "event study" style differences-in-differences plots in Fig. 3C-D, to show how the treatment effect evolved before and after the retraction date. Here we replace the binary treatment times post variable with a series of relative year indicators, each interacted with treatment status. In addition to the reduced form results, we also consider the two-stage-least-squares estimate, where $Treat_Post_{ipt}$ instruments for $Pivot_Size_{ipt}$ (Tables S6-7). As with other researcher-level panel analyses, standard errors are clustered at the researcher level. We next describe specific details of the retraction experiment, including definitions of the treatment and control groups.

S2.6.1 Retractions Analysis

In the retractions analysis, we start with 13,456 papers that were retracted (see SI Section S1.1). Treatment timing is the year of the retraction for each paper. The idea here is that the retraction event for a given paper devalues that research. Scientists who had been drawing on the retracted work may naturally move away from that line of research. The treatment group consists of those authors who had cited the retracted paper at least once in the period between the paper's publication year and the year prior to its retraction but were not themselves an author of the retracted paper. We do not include authors of the retracted papers themselves because these individuals may experience direct effects from the retraction. By contrast, those who had cited the retracted papers can be seen as utilizing work that then appears to have shakier foundations, potentially provoking shifts in the direction of their research. We further analyze the treated group based on how many times the treated authors cited the retracted paper, prior to its retraction. This provides a natural way to consider the intensity of treatment, where authors who were building more regularly on the retracted work may naturally undertake a larger move. Among the treated authors, there are 146,744 authors who cited the retracted paper at least once prior to its retraction and 16,413 authors who cited the retracted paper at least twice.

To build the control group, we consider all authors who cited other publications in the same journal and publication year where the (eventually) retracted paper was published. We then remove from this set any treated author. Among these control authors, we further use coarsened-exact-matching (CEM) so that the control authors match closely to the treatment authors prior to the treatment year in their publication count and rate. There are 146,758 control authors.

S2.6.1 Psychology Analysis

We consider a similar but much smaller natural experiment using replication failures in psychology. Namely, a landmark replication analysis[16] quasi-randomly selected 100 psychology publications from 2008 and tested them for reproducibility. Of the tested papers, 64 had results that failed to replicate, providing the core for our treatment group. Treated authors are those who had repeatedly cited a non-replicating 2008 paper, prior to the replication analysis, but were not themselves an author of the non-replicating paper. As above, we do not include authors of the non-replicating papers themselves because these individuals may experience direct effects from the failure to replicate. By contrast, those who had repeatedly cited the non-replicating papers can be seen as utilizing work that then appears to have shakier foundations, potentially provoking shifts in the direction of their research. There are 843 treated authors. To build the control group, we take all authors who cited other papers in the three psychology journals and publication year chosen for the replication study (but who did not cite any of the papers in the replication analysis). Among these authors, we use coarsened-exact-matching (CEM) so that the control authors match closely to the treatment authors prior to the treatment year in their publication count and rate. Table S7 presents difference-in-differences estimates for this psychology study.

**S2.7 COVID-19 Analyses**

COVID-19 provides another kind of external shock that may encourage pivoting. In contrast to the "push" shock of retractions, where authors may pivot away from research areas that no longer appear reliable, COVID-19 presents an important new object of study, a "pull" shock that may encourage researchers to pivot into this new area. COVID-19 is of additional interest because of its wide societal import and the opportunity to understand how science as a whole pivots to engage a critical new area. Because the shock is to science as a whole, as opposed to specific researchers (as with retraction events), we cannot deploy a natural experiment in the same framework as with retractions. Nonetheless, we can compare researchers who pivot with observationally similar researchers who did not pivot, and we can compare within a given researcher their COVID-19 work with their own other work, to inform how science responds to a critical global shock.

In Fig. 4C, we focus on a subset of COVID authors and non-COVID authors that share similar characteristics. Specifically, these career comparison graphs focus on scientists who first published in 2005. The COVID scientists here are the 6,406 authors among those who first published in 2005 and who published at least one COVID publication in 2020. We then match these authors to a control group of non-COVID scientists that also started their publication career in 2005 and have the same primary level-1 field. For each COVID author, we use a nearest-neighbor match on the number of publications between 2015 and 2019 (sampling without replacement) to construct a control group that also includes 6,406 authors.

We further analyze impact with and without control variables. Fig. 4E considers binned scatter plots, comparing COVID and non-COVID research, without control variables. In Fig. 4G-H and Figs. 4J-K, we split the COVID and non-COVID samples into groups based on the median prior impact, career age, team size, and number of new collaborators, and whether the paper was connected to a funding source.

For the wide-ranging multivariate regression results presented in Fig. 4K, we include additional controls. The additional controls include fixed effects for average prior impact groups, author age groups, team size, the number of new collaborators, and an indicator variable for whether the paper was funded. In Fig. S18, we include individual fixed effects, or in other words control for all fixed characteristics associated with each author. In this individual fixed effect analysis, we are considering researchers who produce both COVID papers and non-COVID papers in 2020, thus allowing comparison of outcomes and pivoting within individuals who respond to the pandemic and comparing outcomes within their contemporaneous body of works.

## S3 Additional Analyses

### S3.1 High Pivot Cases

Cases of high pivots might potentially be an artifact of name disambiguation, where two different people are conjoined into one record but work in different areas. One test for this is to hand-check high pivot cases, comparing database results against public CVs. To proceed, we took a random sample of 10 authors who produce a paper with a pivot score >0.95 in the year 2020. Of these 10 authors, 5 were randomly chosen from authors with the 200 most common names, and 5 were randomly chosen from authors with uncommon names. For each author, we then took their very high pivot paper (10 papers) as well as all other papers associated with that author in the database that were published over the prior three years (totaling another 148 papers). We then hand checked every paper in the database associated with these authors against the authors' own CVs, personal websites, Google Scholar profile, Scopus page, or PubMed page (depending on what source was available for a given author).

The results of this manual verification were as follows. First, for the 10 very high pivot papers, we found each paper on the authors' own CVs/ websites / Google scholar etc. profiles. Thus, all the high pivot papers appear correctly assigned to these authors. Second, examining the prior works of these authors, for 9 of the 10 authors we located 100% of their prior papers in Dimensions on the authors' public profiles. For the remaining author, who is located in China, we could verify 27 papers (80%) on the author's Scopus and PubMed profiles, where the other 7 papers in Dimensions were in Chinese-language journals; these match on name and field of the author but are not listed in the extant English-language profiles, so we could not confirm that Dimensions was correct, or incorrect, for these 7 papers. In sum, this manual verification exercise suggest that all the high pivot papers were correctly assigned to these authors, and we could confirm 96% of the works of these authors are also correctly assigned, while the other 4% of papers had clear matching characteristics but we could not verify the match against other profiles. The very high success rates matching these authors' works by hand gives further confidence that name disambiguation is sufficiently accurate. A spreadsheet detailing the hand curation exercise for each of the 158 papers analyzed is available from the authors upon request.

Another approach, which can be scaled across the data, is to analyze generally common names and rare names. The idea is that researchers with common names may pose greater challenges for name disambiguation, which in our case would be revealed as showing larger apparent pivot sizes. To proceed, we take all papers published in 2010. We then plot a binscatter relating mean pivot size to surname frequency (Fig. S24). While we see some variation, overall the relationship is quite flat, with the most common surnames showing similar pivot sizes as seen among relatively uncommon surnames. We further consider the pivot penalty relationship, separately for both individuals with the most common surnames and, separately, among other authors (Fig. S24). We see that the pivot penalty appears in both groups. Overall, these additional analyses further increase confidence in name disambiguation and the robustness of the findings.

### S3.2 Outlier Fields

Table S1 Panel A indicates that a large majority (93.5%) of the 153 fields with at least 20 papers show a negative correlation between pivot size and impact. Here we investigate the 6.5% of fields (10 fields) that do not show this negative correlation.

An initial observation is that these outlier fields are relatively small. Although these 10 fields are 6.5% of fields, they collectively incorporate only 0.18% of papers. Table S8 lists these 10 fields, together with their observation counts, and the slope and its standard error when relating pivot size and impact. Of the ten fields, five show a statistically significant relationship. One of these "Other Built Environment and Design" has only 87 papers.

Another important observation is that the outlier fields have characteristics that may make the pivot measure less salient in these specific contexts. Specifically, there are notable field commonalities. First, the only field with a highly statistically significant positive relationship is "Visual Arts and Crafts," and the field Art Theory and Criticism also exhibits a positive (but non-significant) relationship. One interpretation is that art-oriented fields privilege pivoting, but these are also fields where books are a main avenue of output and references, and thus our journal-reference pivot measure may be less salient in defining reference and pivoting behavior. Second, the largest three outlier fields in size are all computer-science related fields. These are relatively small subfields of computer science; the larger computer science fields exhibit the usual pivot penalty. Further, computer scientists rely heavily on conference proceedings as key venues for their work. In publication databases each conference proceeding acts as a different journal in each year. This may lead to apparent high pivots as these "journals" come and go, making the pivot measure less salient, while at the same time these conference proceedings can be associated with high impact new work in computer science.

One may further consider higher level groupings of fields to see if there are any important contrasting areas of research. In Figure S21, we categorize the L1 fields into 7 higher level groupings, mapping each field into one of medical sciences, life sciences, physical sciences, engineering, social sciences, humanities, and other. As can be seen in the figure, there is some heterogeneity, but these different areas of research all show substantial, negative relationships between pivot size and impact.

Overall, the small minority of fields with the contrasting relationship are those that have relatively low numbers of publications and those where journal-based pivot measures may be less effective in capturing reference and pivoting behavior.

**S3.3 Pivoting and Field Switching in Science**

The pivot size measurement framework quantities shifts in research direction on a [0, 1] interval, allowing assessment of research shifts in a continuous manner. The method can also be applied using alternative encodings of research areas. In science, we have alternatively studied pivots based on journals as well as field encodings. In patenting, we have alternatively studied pivots using various levels of detail in hierarchical technology class encodings.

One may also be interested in relating the magnitude of pivot sizes, and the pivot penalty, to the case where a scientist switches to a new field. Specifically, one can examine the relationship between the pivot measure (examining the references in a paper) with a binary measure of switching fields (where field codes are assigned to a paper). Figure S22 shows that when an individual engages a new field, that paper will be associated with a large average increase in pivot size. The top row shows the results for switching to a new L0 field (there are 22 L0 fields), and the bottom row shows the results for switching to a new L1 field (there are 154 L1 fields).

Switching to a new L0 field code is associated with moving from a pivot size of approximately 0.5 on average to approximately 0.7 on average. Considering the pivot penalty, such an increase in average pivot size is associated with a hit rate decline by approximately 2 percentage points in recent periods (see Fig. 2). In short, switching to a new field is associated with an approximate 40% drop in the probability of writing a high-impact paper. Note that the quality of the paper-level field encodings in large bibliometric databases is the subject of debate[17,18], suggesting potential noise in measuring field switches, which may in turn attenuate its relationship with pivot size and impact.

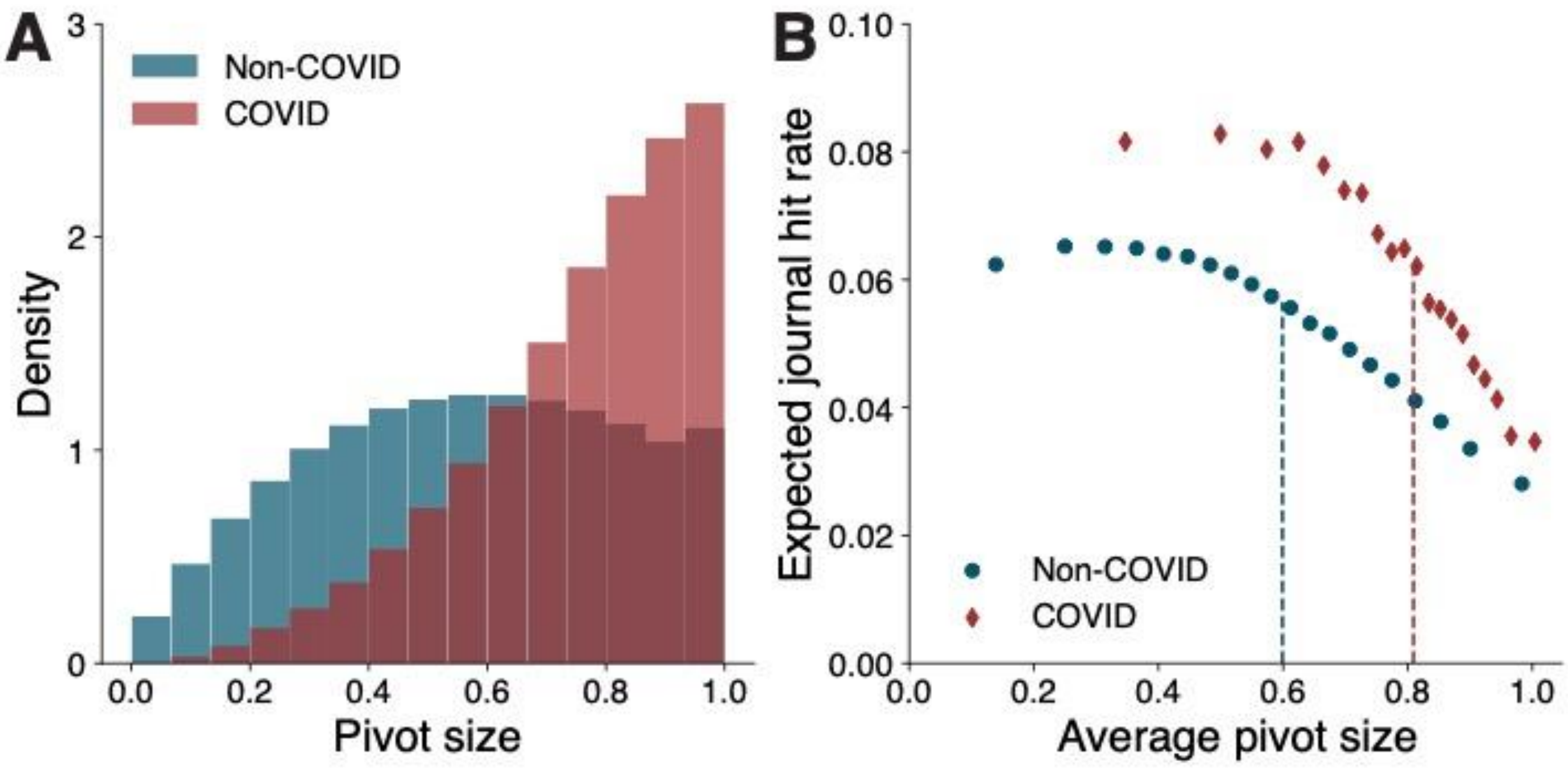

**Figure S1: Quantifying pivot size using an author's full publication record.** In the main text, we measure pivot size comparing the author's focal paper with that author's prior three years of work. Here we examine pivot size using the entire history of that author's work. **(A)** The large shift in pivot size for COVID papers is evident when pivot size is measured by comparing 2020 papers to all past work. This shift is comparable to Fig. 1B, where pivot size is measured using only papers published in the prior 3 years. **(B)** The negative relationship between pivot size and impact is similar in slope when using the full career pivot metric here or the 3-year metric as shown in Fig. 4E.

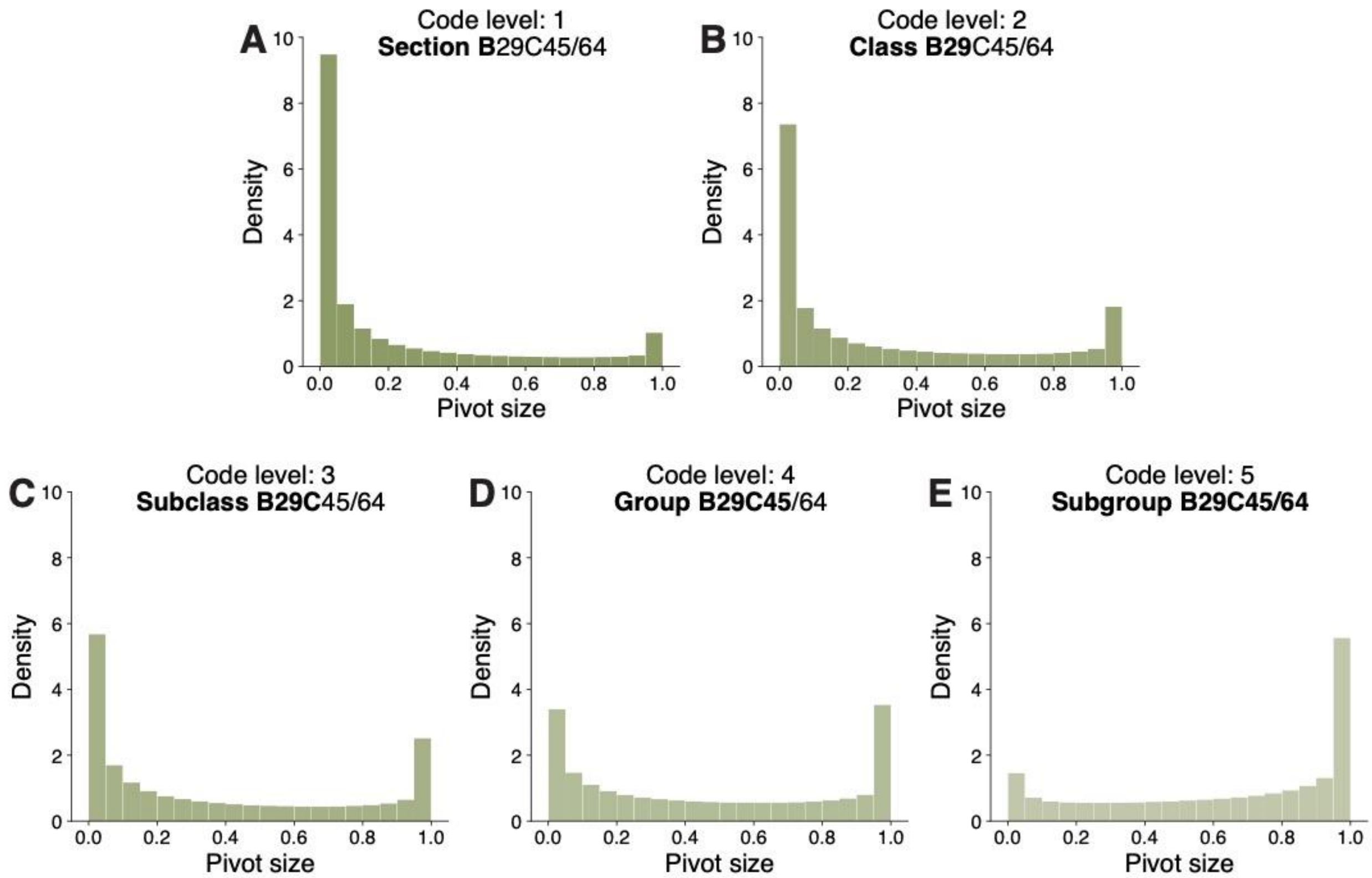

**Figure S2: Quantifying pivot size using various levels of patent technology classification.** For historical patents granted from 1975-2015, the pivot size distribution is bimodal, with more weight on pivots of size zero and one. The average pivot size increases as the definition of technology class used to calculate pivoting narrows. The available levels of technology class are: (**A**) 9 sections (e.g., "B"), (**B**) 128 classes (e.g., "B29"), (**C**) 662 subclasses (e.g., "B29C"), (**D**) 9,987 groups (e.g., "B29C45"), and (**E**) 210,347 subgroups (e.g., "B29C45/64"). The main analysis in Figures 1 and 2 use level-4 groups to define pivot size.

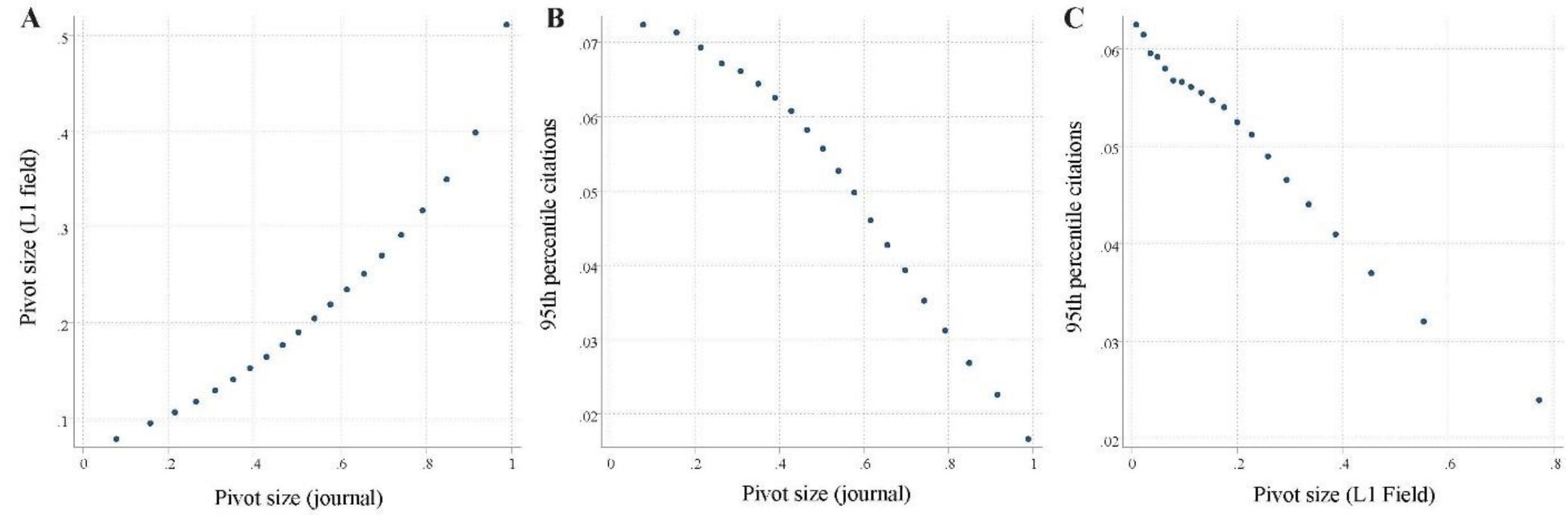

**Figure S3: Quantifying pivot size and pivot penalty using L1 field codes.** In the main analyses, we use journals cited in reference lists to build the referencing vectors and calculate pivots. Here we consider pivots using the L1 fields of the referenced papers, rather than the papers' journals. This is a coarser approach, as there are 154 L1 fields, as opposed to tens of thousands of journals. (**A**) Presents the relationship between pivot size using L1 fields versus journals. We see a positive relationship. We also see a narrowing of the pivot size distribution when using L1 fields, indicating that researchers naturally shift less when the measure uses wider encodings for areas of knowledge. (**B**) and (**C**) present bin scatters. We see the pivot penalty is robust to using the field encoding. Again the distribution of pivot size is substantially condensed, with only a small share of papers having pivot sizes above 0.5. See Figs. S2 and S4 for similar analyses for patents, calculating pivot size using coarser and finer technology classifications.

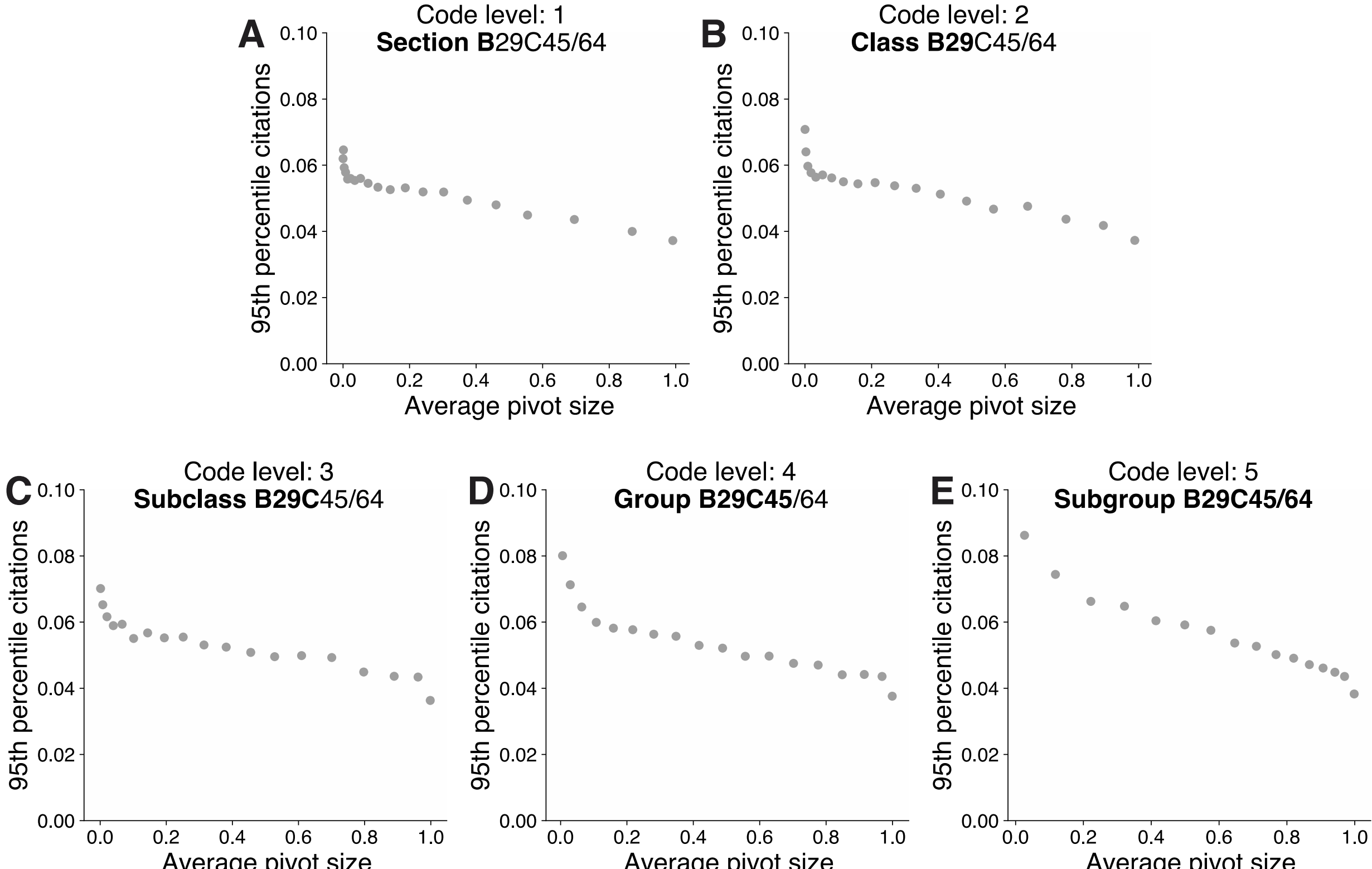

**Figure S4: The pivot penalty with various technology levels.** The probability of being a highly cited patent is decreasing in pivot size for all technology code levels used to define pivoting. The difference in impact between the highest and lowest pivot size is (**A**) smallest when using broad level-1 classes and (**E**) largest when using narrow level-5 subgroups.

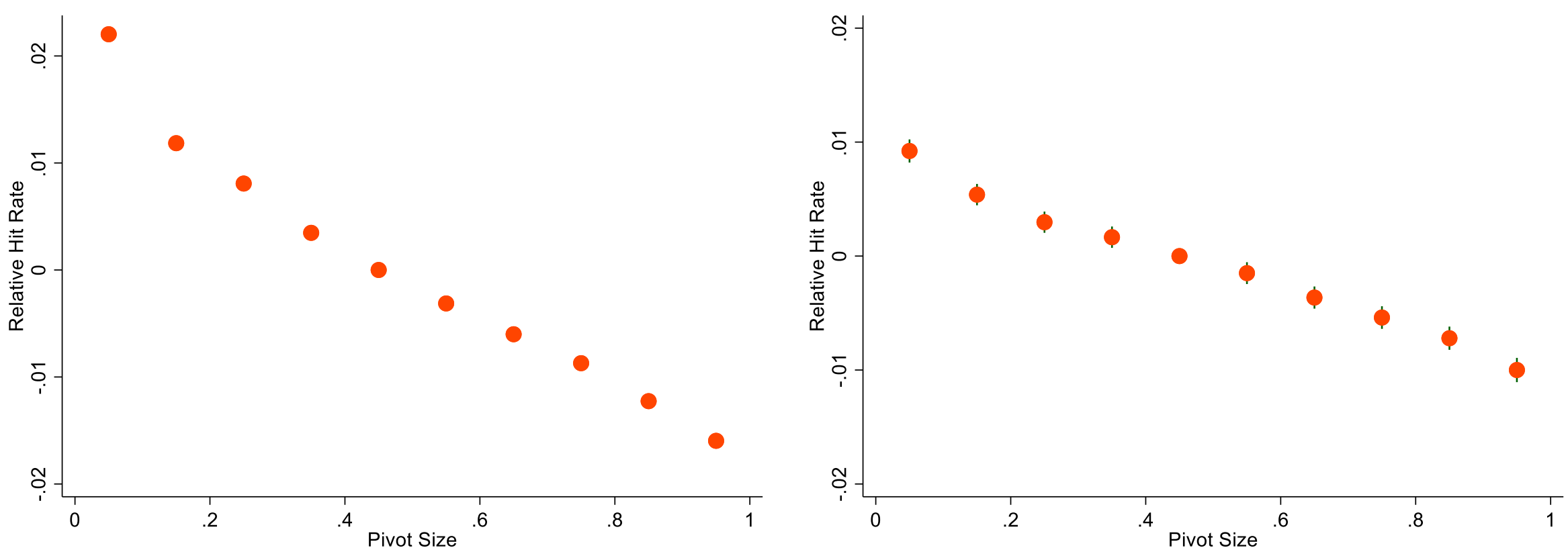

**Figure S5: The pivot penalty in science over time, within individual researchers.** We divide the data into two periods, 1986-2000, and 2001-2015. In each period, we run regressions with individual fixed effects. The relationship between hit rates and pivot size is estimated non-parametrically, with fixed effects for different ranges of pivot size. The figures present the coefficient for each pivot size group, with indicated 95% confidence intervals. The slope of the pivot penalty is increasing over time when looking within individual researchers. The recent period (left panel) shows a monotonic decrease in hit rate with pivot size, within the body of work of individual researchers (confidence intervals are too small to be seen). The earlier period (right panel) similarly shows a monotonic decrease in hit rate with pivot size, but the slope of the relationship is shallower. Overall, we see an increasingly steep pivot penalty with time.

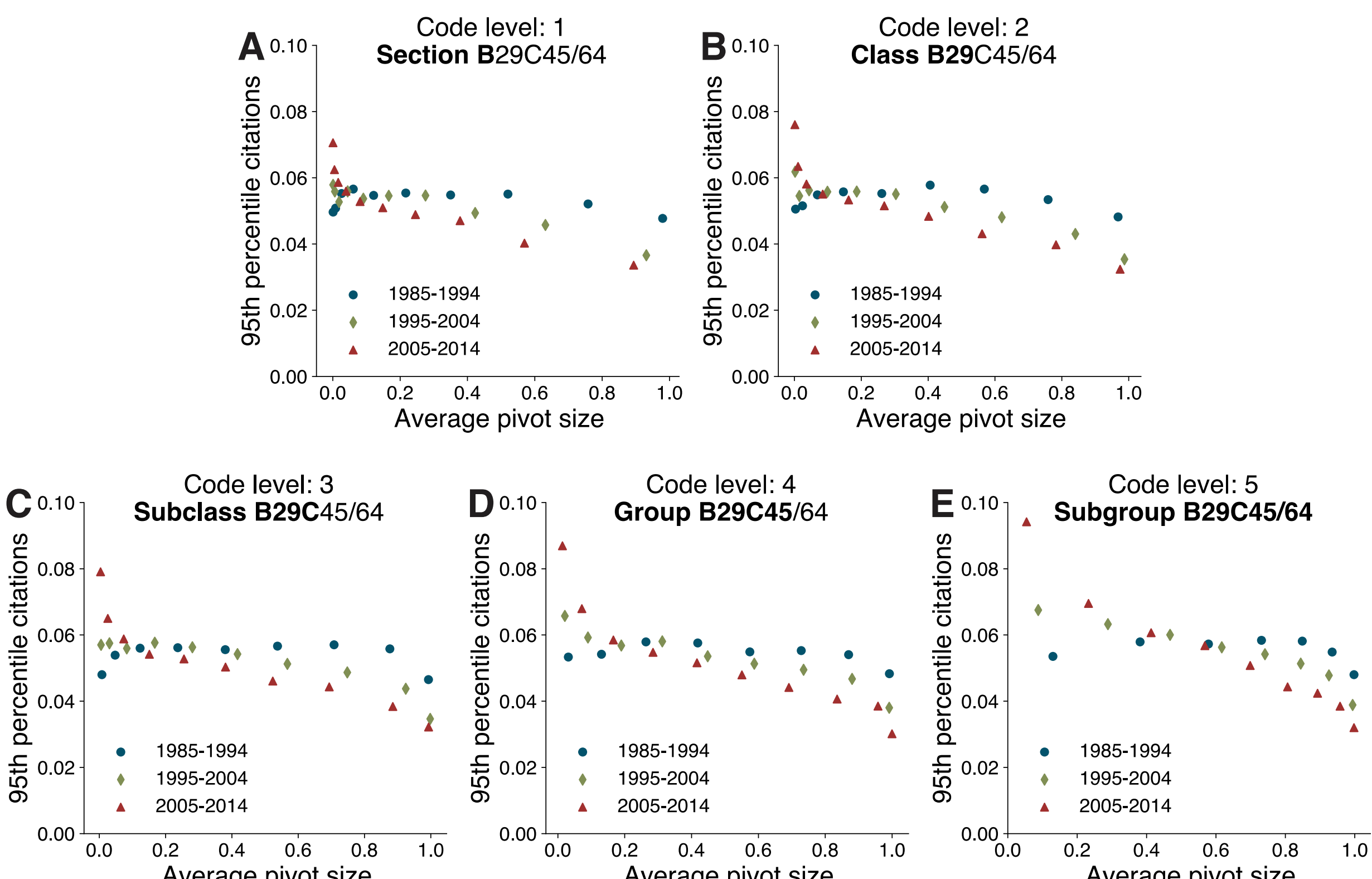

**Figure S6: The pivot penalty over time with various technology levels.** The slope of the pivot penalty is increasing over time, regardless of which level of technology code is used to define pivot size. The increase in slope over time is (**A**) smallest when using broad level-1 sections and (**E**) largest when using narrow level-5 subgroups.

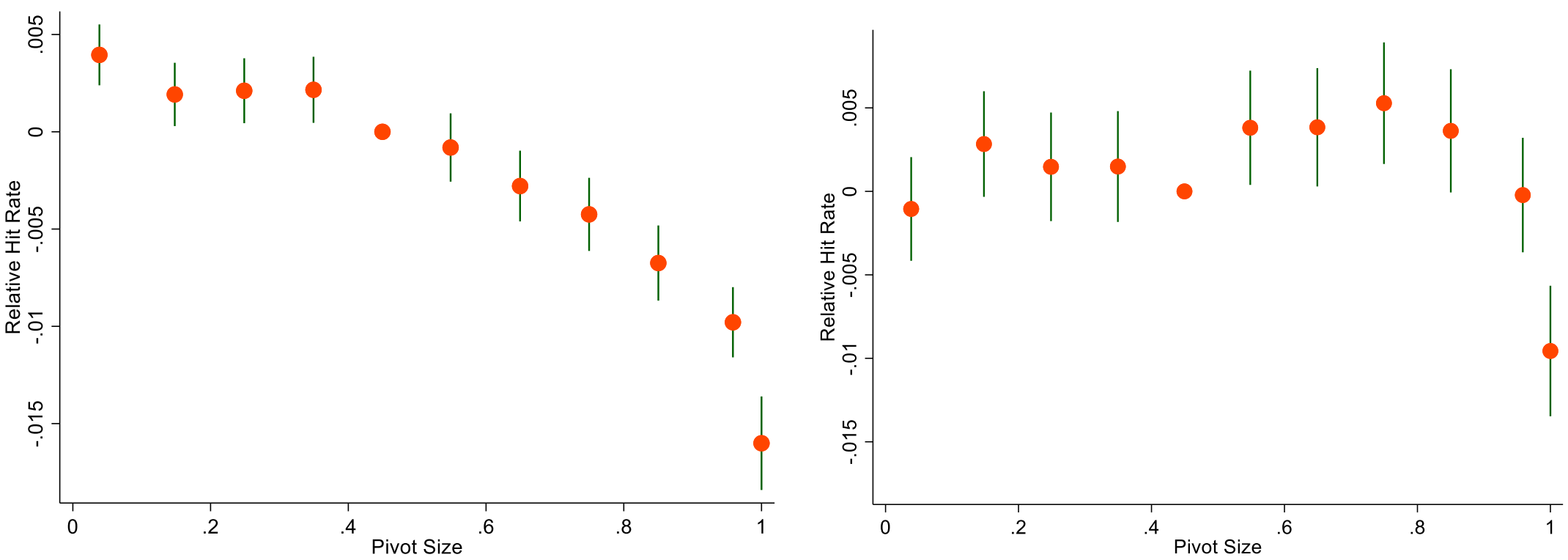

**Figure S7: The pivot penalty in invention over time, within individual researchers.** We divide the data into two periods, 1986-2000, and 2001-2015. In each period, we run regressions with individual fixed effects. The relationship between hit rates and pivot size is estimated non-parametrically, with fixed effects for different ranges of pivot size. The figures present the coefficient for each pivot size group, with indicated 95% confidence intervals. The slope of the pivot penalty is increasing over time when looking within individual researchers. The recent period (left panel) shows a monotonic decrease in hit rate with pivot size, within the body of work of individual researchers. The earlier period (right panel) has noisier estimates, with a flatter relationship to pivot size and potential non-monotonicity, but where high pivots face large impact penalties.

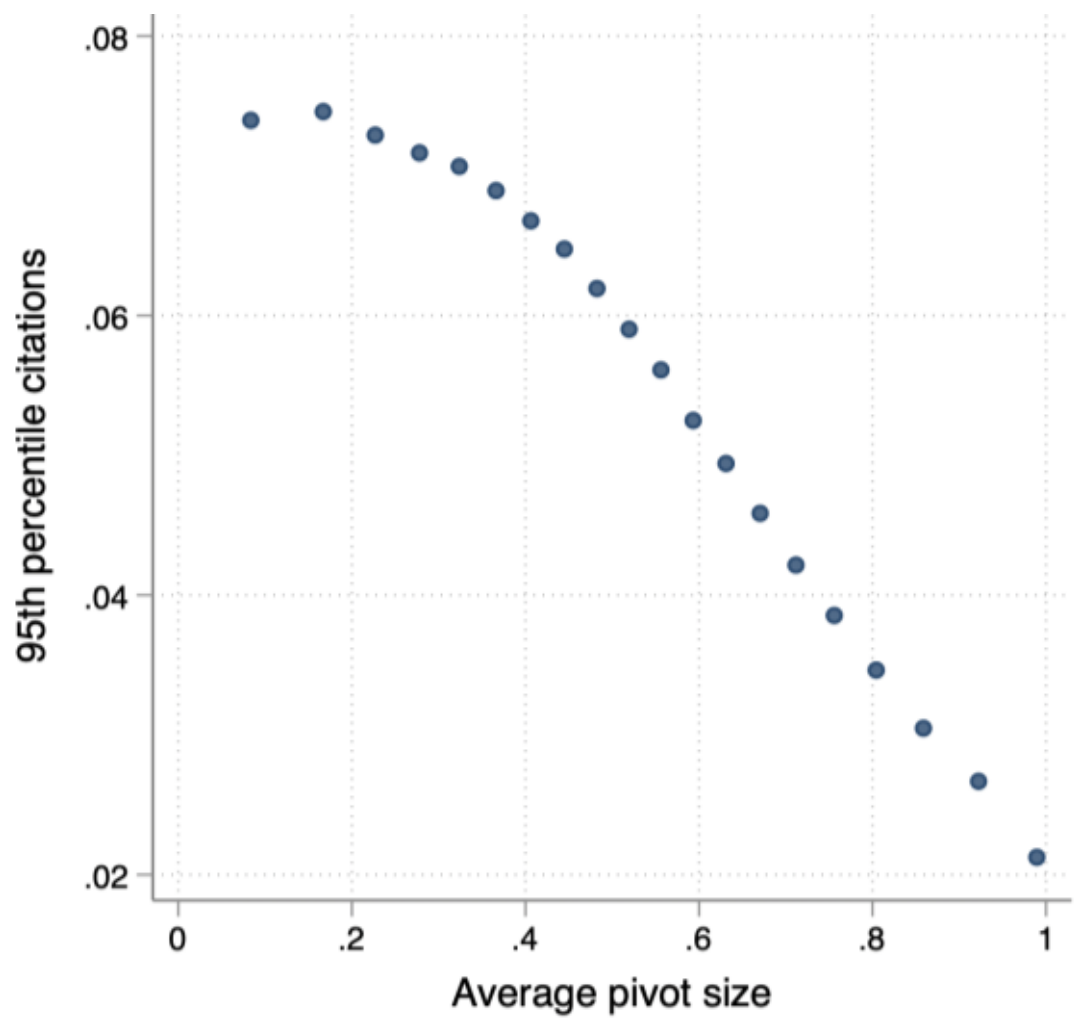

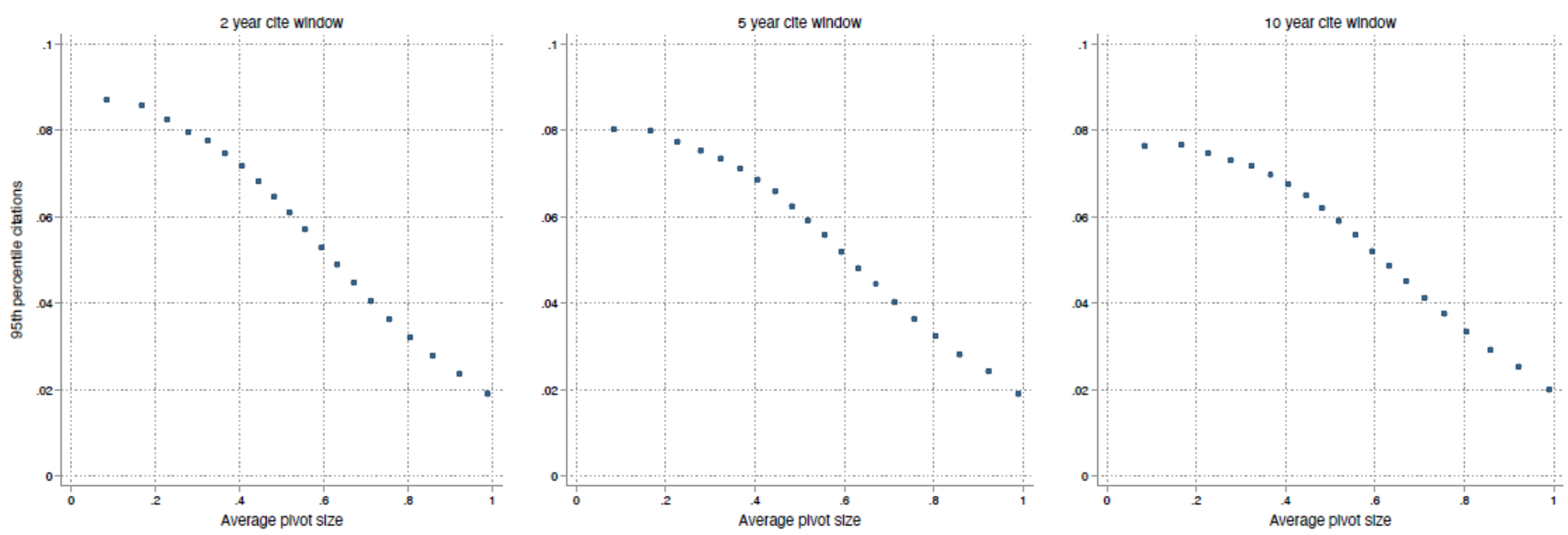

**Figure S8: The pivot penalty over alternative time horizons.** The hit rate measures in the paper normalize impact by field and publication year, providing one means for confronting different time horizons for citations from different publication years. Alternatively, one can count citations received only over a fixed window of time after the publication year. (**Top**) The pivot penalty result using the baseline hit rate measure. (**Bottom**) Hit rates with citations counted over a fixed, five-year citation window for all papers. Hit rates are computed using citations received by each paper over, alternatively, 2 year, 5, year, and 10 year forward windows, as indicated. The pivot penalty is robust and extremely similar using all of these alternatives.

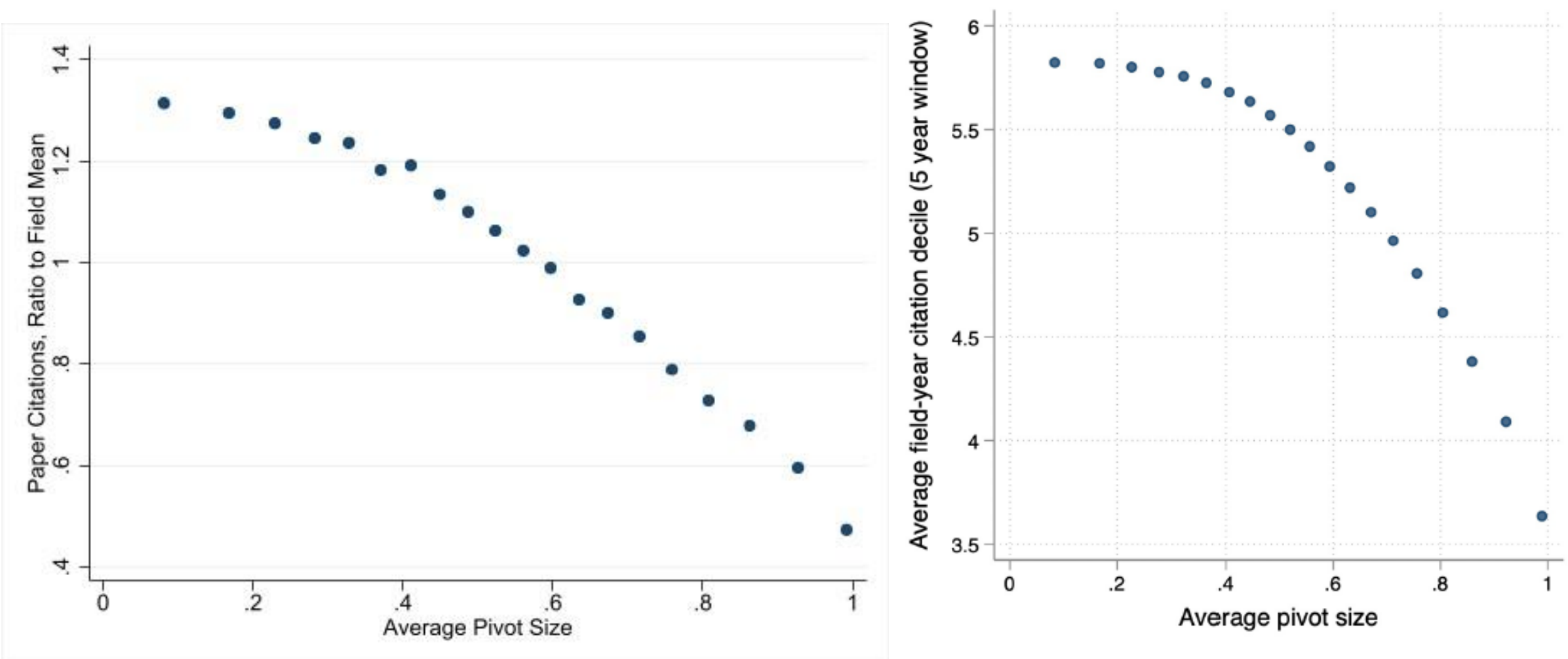

**Figure S9: The pivot penalty with smoother citation measures.** In addition to binary measures of impact, one can consider more continuous measures. In (**A**) we normalize each paper's citation count by the mean citation for papers in that field and publication year. Citations are approximately 30% above the field mean for low pivot papers on average and 55% below the field mean for the highest pivot papers on average. In (**B**) we normalize each paper's citations by its percentile in the citation distribution for all papers published in the same field and year. The pivot penalty is also robust to this measure of impact.

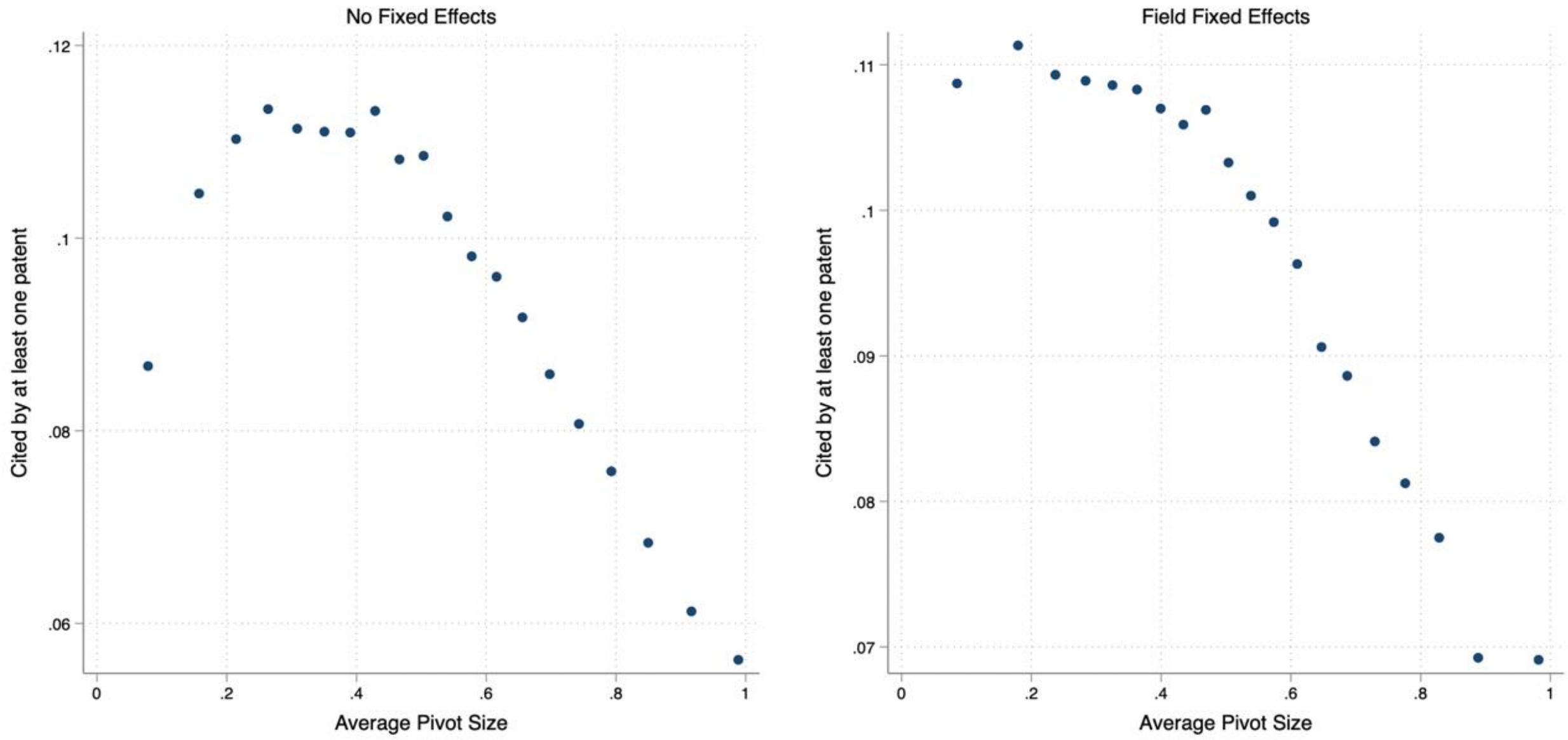

**Figure S10: Patent references to papers.** The probability that an academic paper is referenced by at least one patent declines at larger pivot sizes. The data considers 37 million papers published from 1970-2019. The left panel considers the raw data pattern, with no controls, and indicates non-monotonicity at lower pivot sizes. The right panel considers the relationship net of L1-field fixed effects, which accounts for the fact that some fields (e.g., astronomy) are far less likely to be referenced in patents than others (e.g., nanotechnology). As seen in the figure, controlling for field largely eliminates the non-monotonicity.

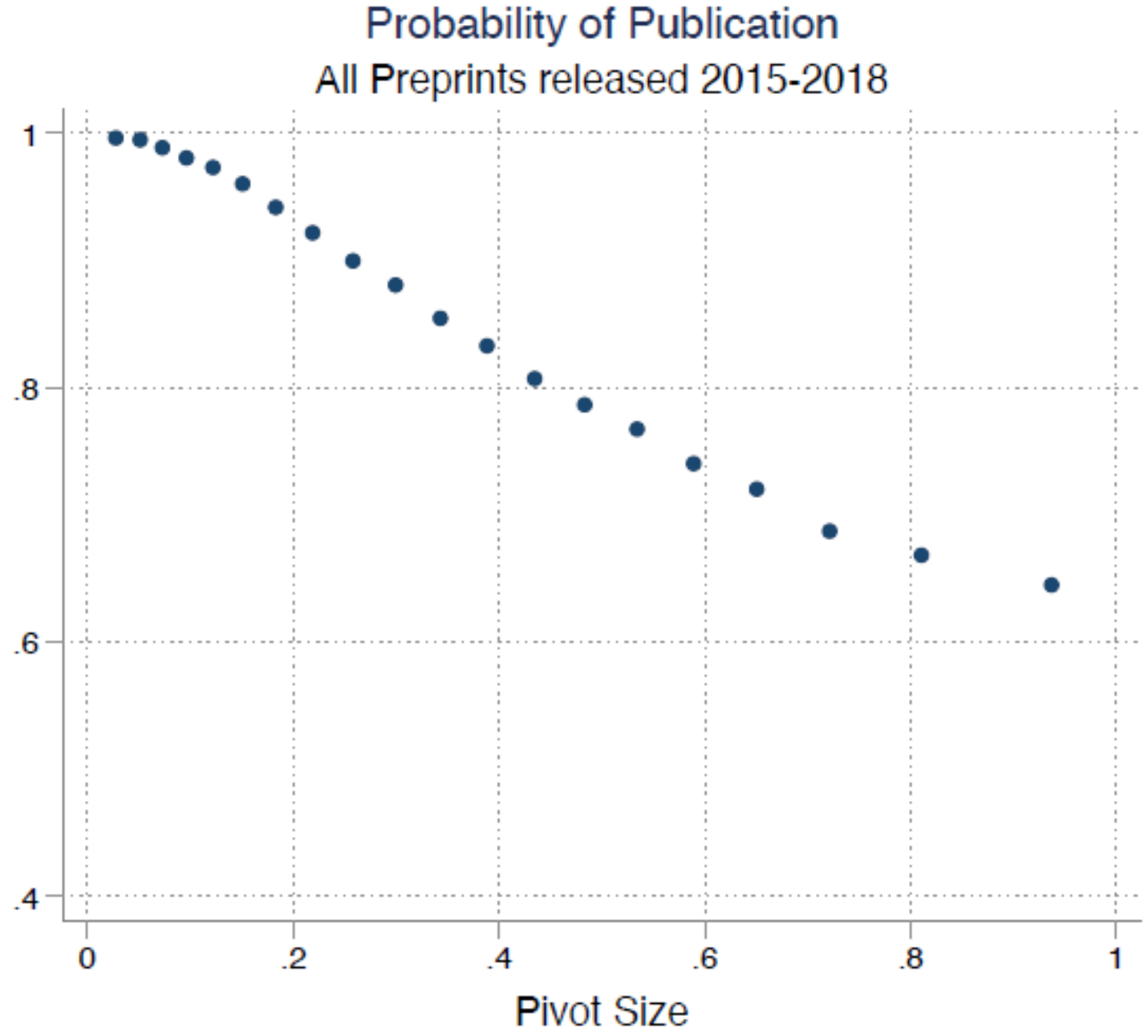

**Figure S11: Successful publication.** This figure analyzes all working papers released from 2015-2018, using preprint databases such as arXiv and SSRN. For each working paper, we examine whether it has been published within a five-year window from its preprint date. Virtually all low pivot size papers are published. But publication success declines smoothly with pivot size, with less than two-thirds of the highest-pivot papers successfully publishing. This monotonic decline in publication success provides a further, reinforcing dimension of the pivot penalty. See Section S2.2.2 for further discussion.

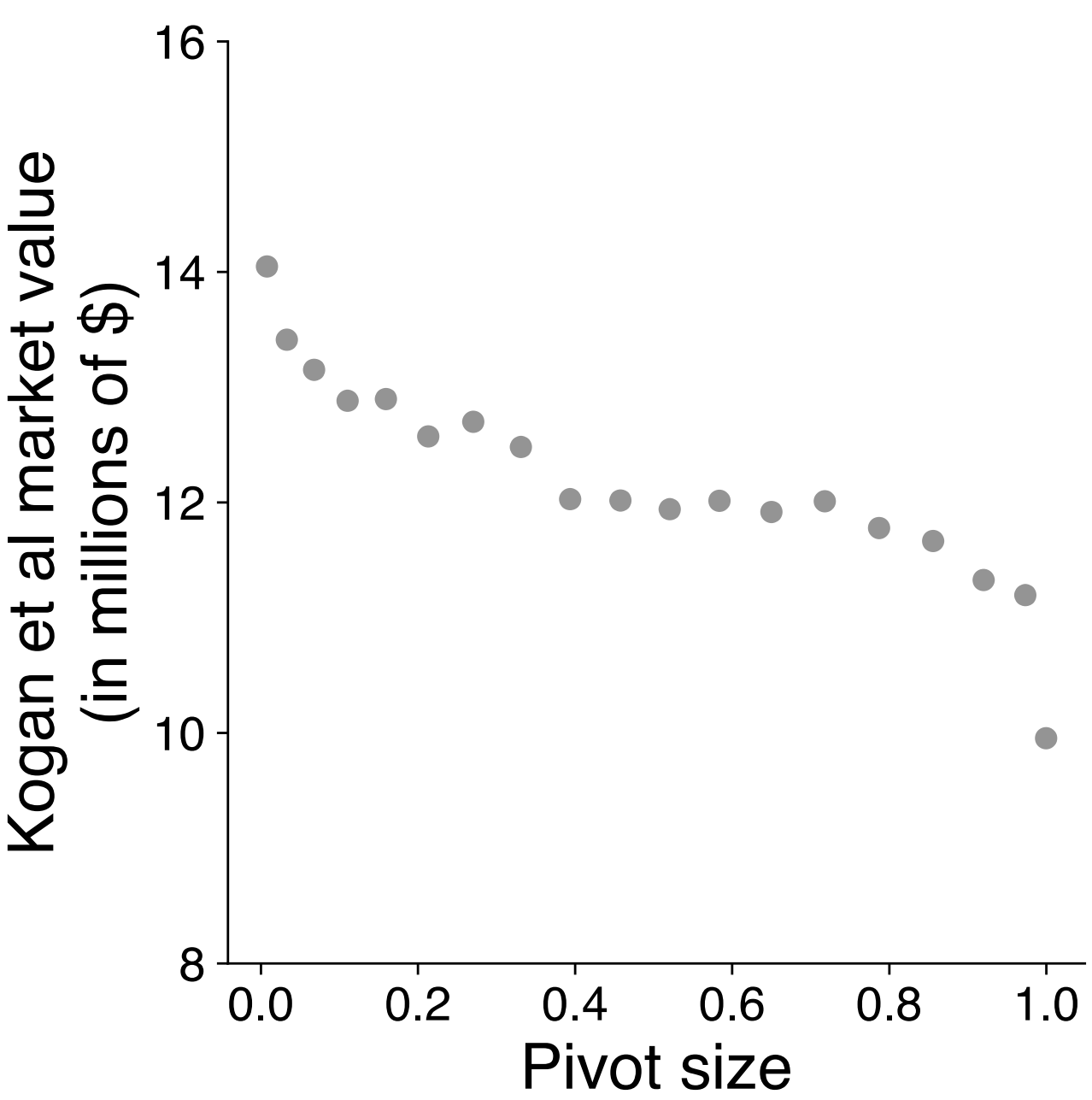

**Figure S12: Patent market value.** The estimated market value of patents is decreasing in average pivot size. Market value is estimated using changes in stock prices around the announcement of patent grants for public companies. The sample is 802,599 patents published between 1980 and 2015 that were granted to public corporations. Market valuations as calculated by Kogan et al (2017).[7]

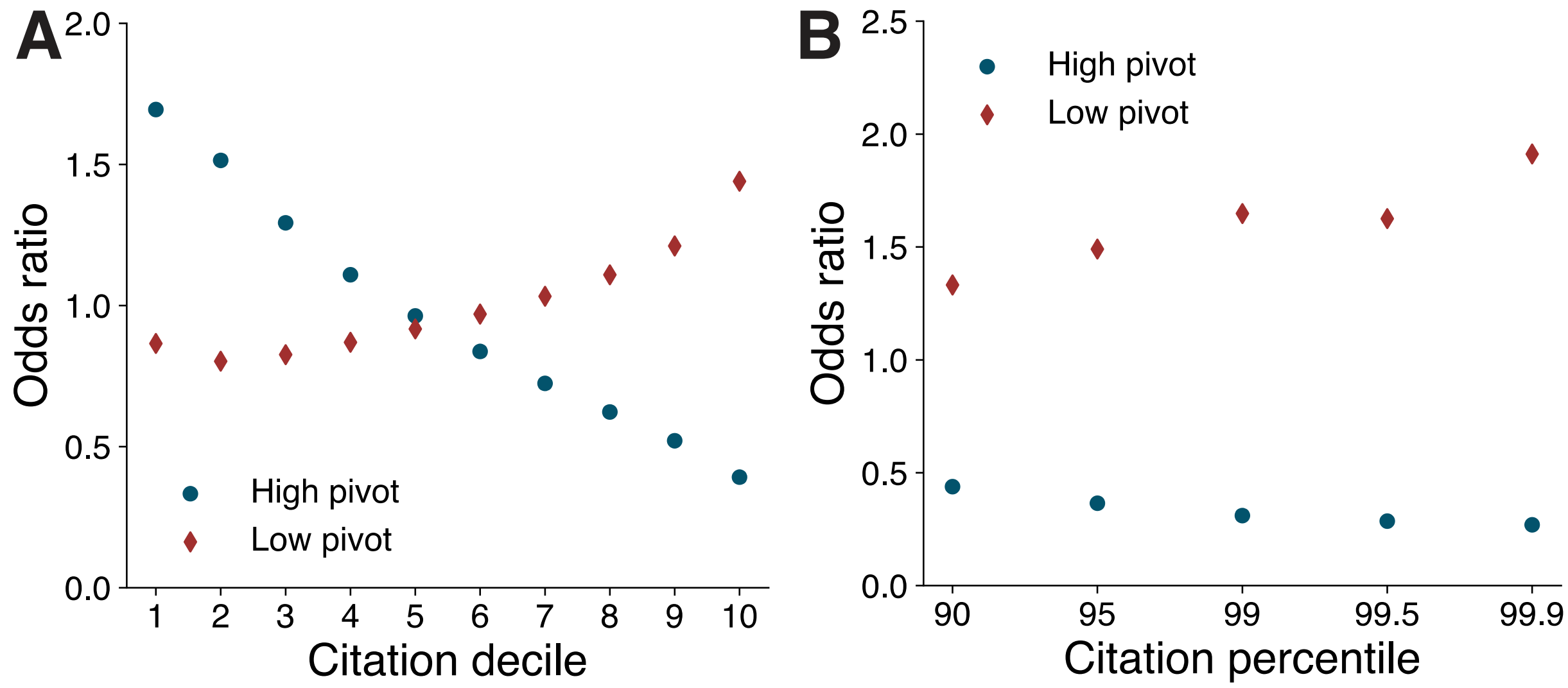

**Figure S13: Probability of high and low pivot across citation distribution.** **(A)** The x-axis groups all papers into deciles by the number of citations within year and L0 field. The y-axis reports the odds ratio that a low and high pivot paper will be found in that citation bin. The low (high) pivot odds ratio is calculated as the share of papers in each citation decile that are in the lowest (highest) decile of pivot size divided by the share of all papers in that decile. Papers in the lowest citation decile are almost twice as likely to be high pivot papers than low pivot, while papers in the highest citation decile are almost three times as likely to be low pivot papers than high pivot. **(B)** The x-axis groups all papers into upper percentiles by the number of citations within year and L0 field. The y-axis reports the odds ratio that a low and high pivot paper will be found in that citation bin. Low pivot papers are 3-7 times more likely than high pivot papers to surpass the highest thresholds of impact between the $90^{th}$ and the $99.9^{th}$ percentile of citations.

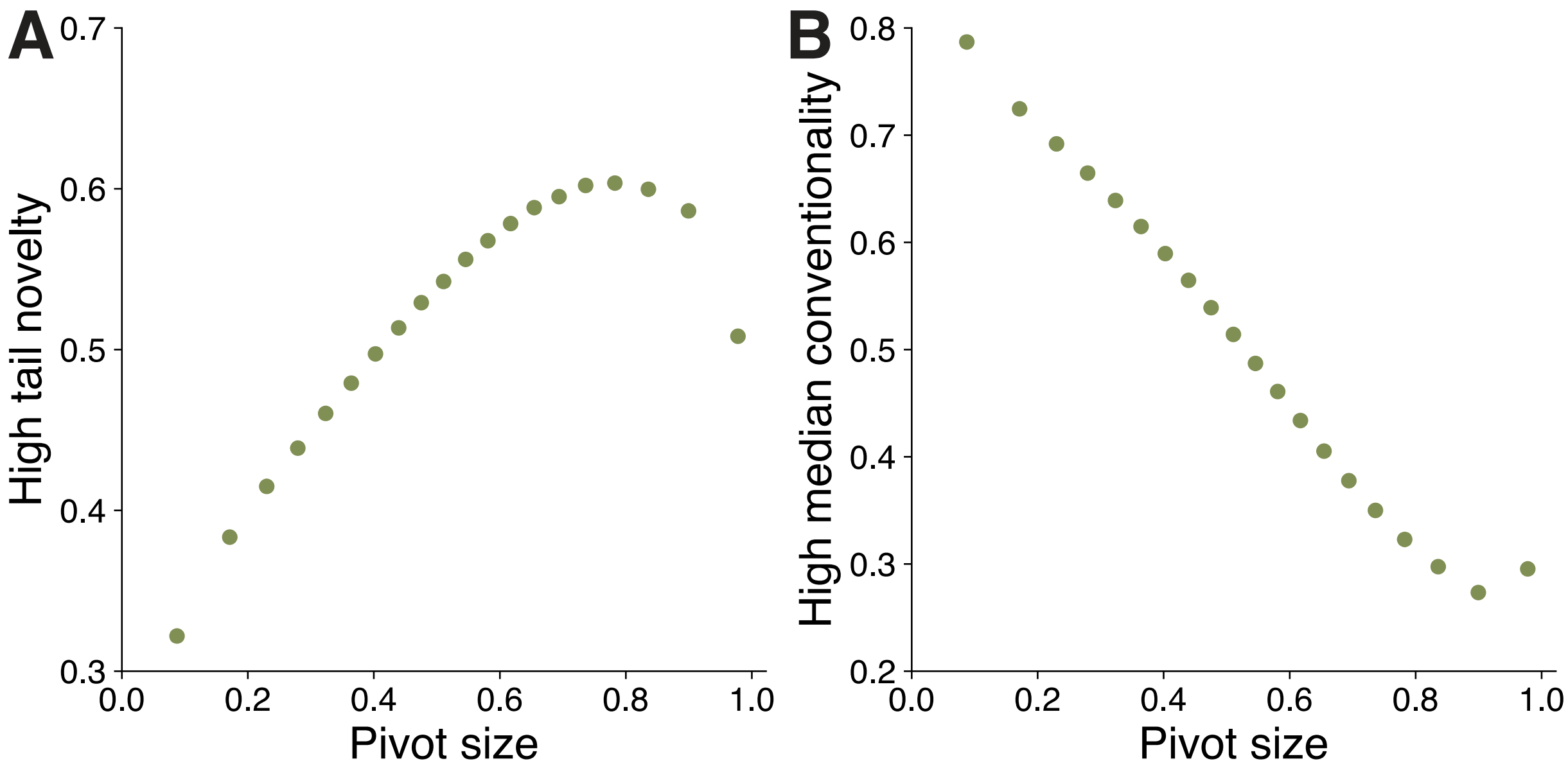

**Figure S14: Novelty, conventionality and pivot size.** **(A)** Novelty is increasing with pivot size while **(B)** conventionality decreases. Measures are calculated using combinations of references in new academic papers[12]. A researcher who is pivoting not only does something new personally, but also tends to combine prior knowledge in a way that is unusual in science. At the same time, high pivots are associated with distinctly low conventionality, consistent with a weaker grounding in conventional domain knowledge.

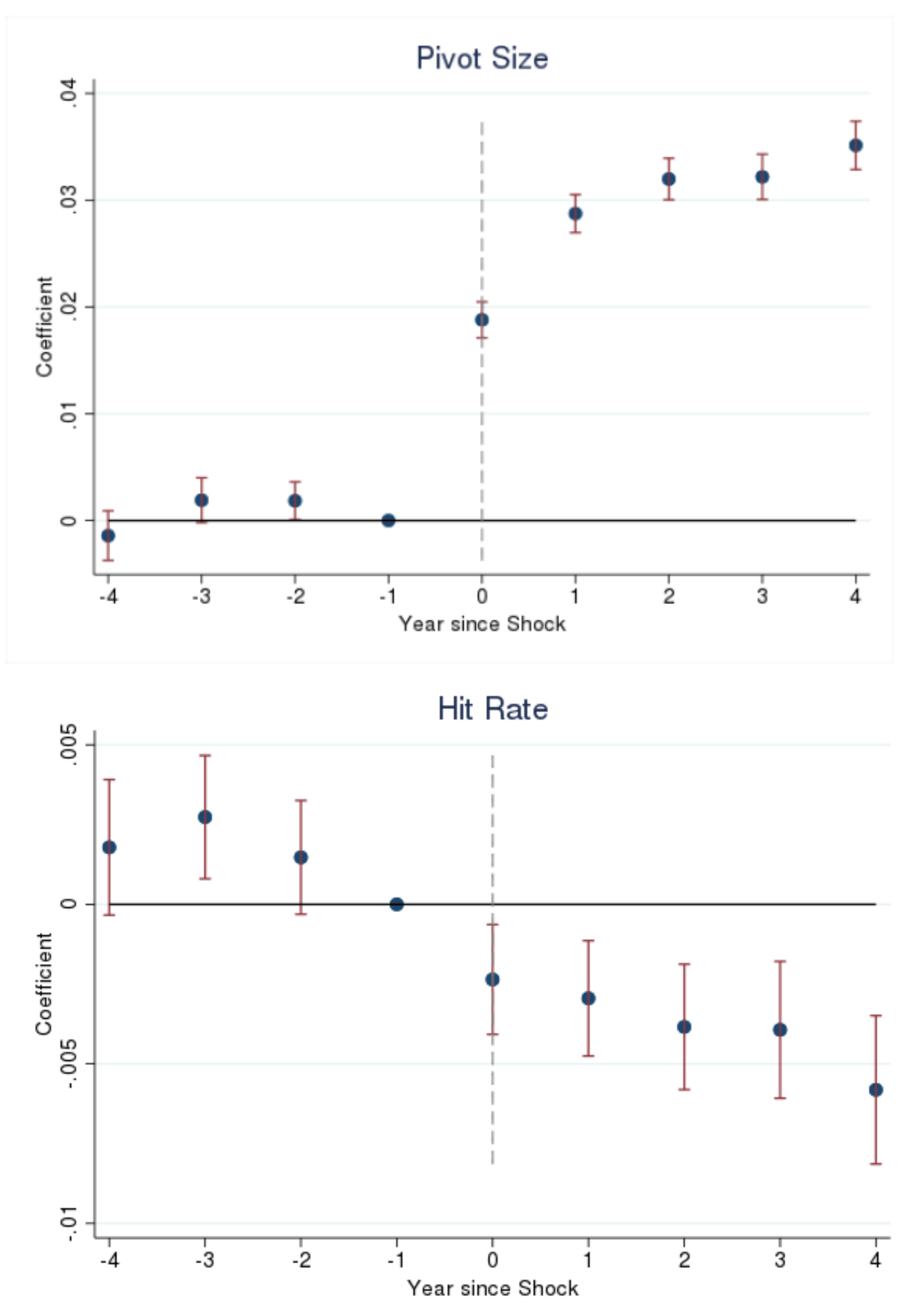

**Figure S15: Difference-in-Differences, 1+ reference group** Figure 3C-D present event study plots for pivot size and hit rate, defining the treated group as those who referenced the retracted paper multiple times. Here we present event study plots for (**A**) pivot size and (**B**) hit rate but now using the broader set of researchers who reference a retracted paper one or more times. We see similar results as in Figure 3, with an increase in pivot size and a decrease in pivot size after the retraction event.

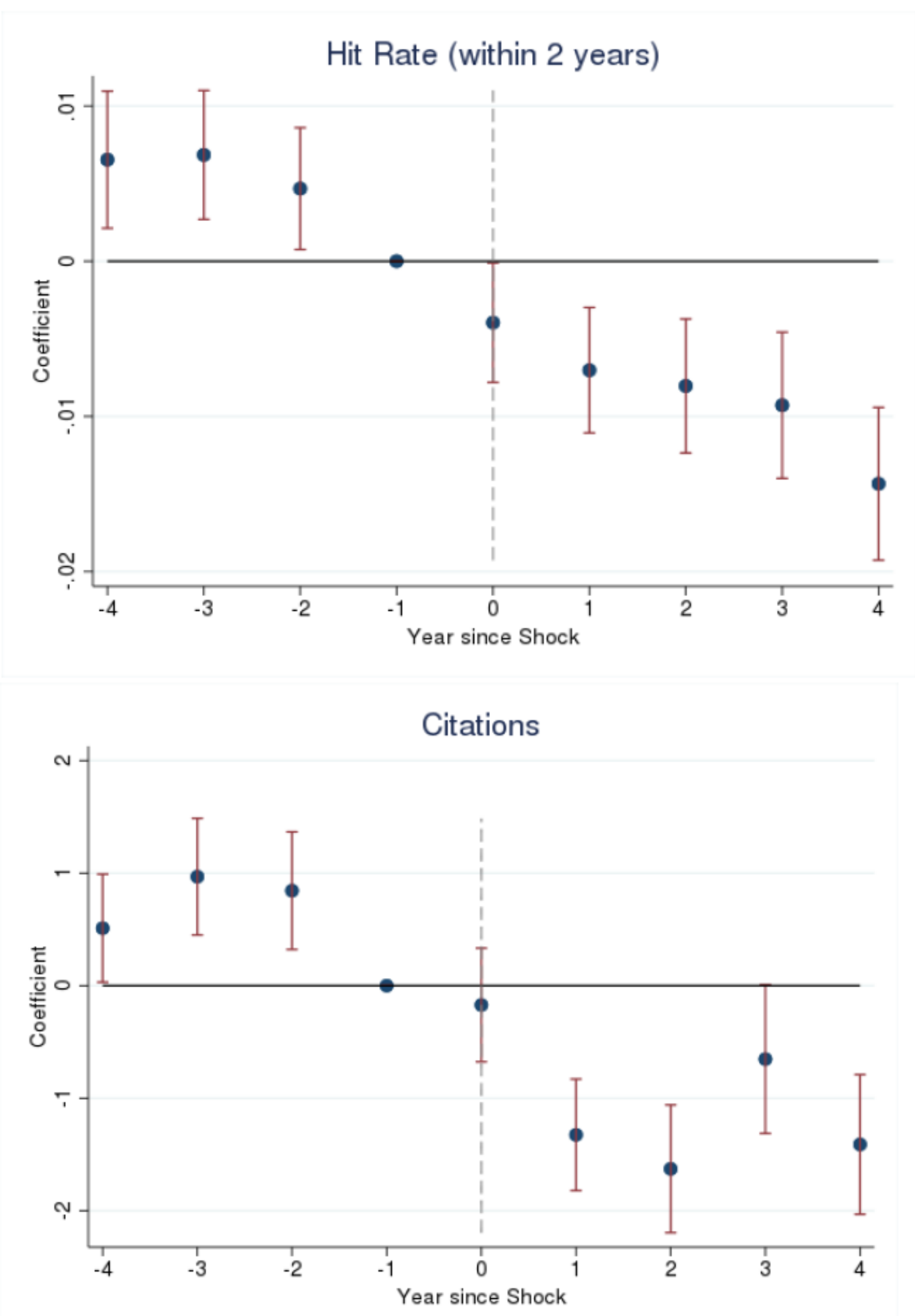

**Figure S16: Difference-in-Differences, alternative impact measures** Fig. 3D presents an event study plots where the hit rate is measured using the whole forward citation window after papers' publications. Here we consider alternative impact measures. In the top panel, the hit rate is measured using the first two years of citations after publication. In the bottom panel, the outcome is a direct citation count over two years after publication. To the extent that retraction events devalue areas of knowledge, these events may reduce references to pre-period works related to the retracted paper. This would cause pre-period paper impact to fall among treated authors, resulting in a conservative bias by making it harder to detect an impact decline for post-period works. Looking at the first two years would mute any effect of the retraction event on citations to the earlier stream. These analyses further support the findings, in addition to acting as robustness tests to time windows more generally.

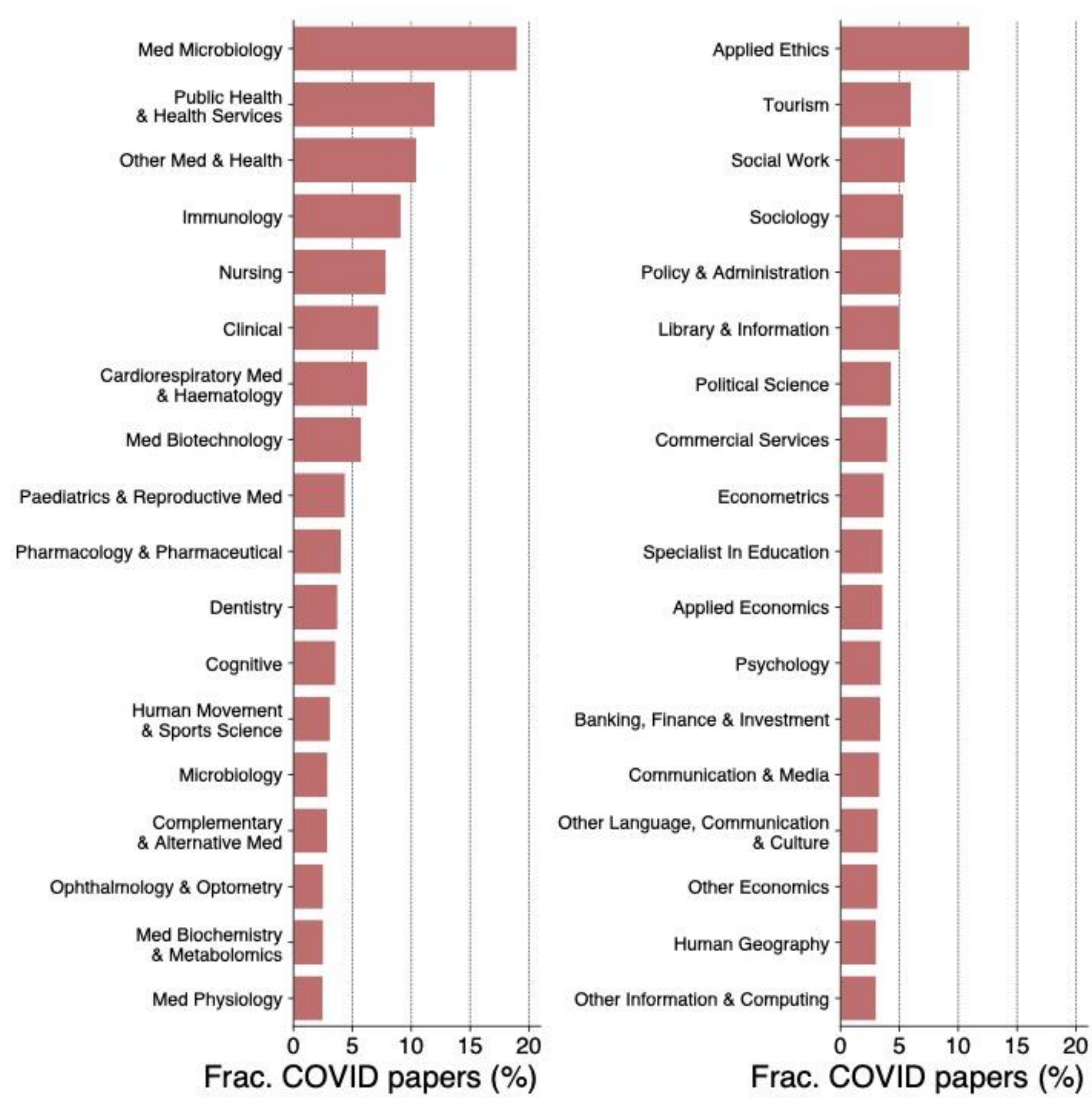

**Figure S17: COVID share by subfield**. This figure reports COVID-19 papers as a fraction of all 2020 publications in specific level-1 fields. Presented here are the 20 medical and 20 non-medical level-1 fields that have the highest fraction of COVID papers.

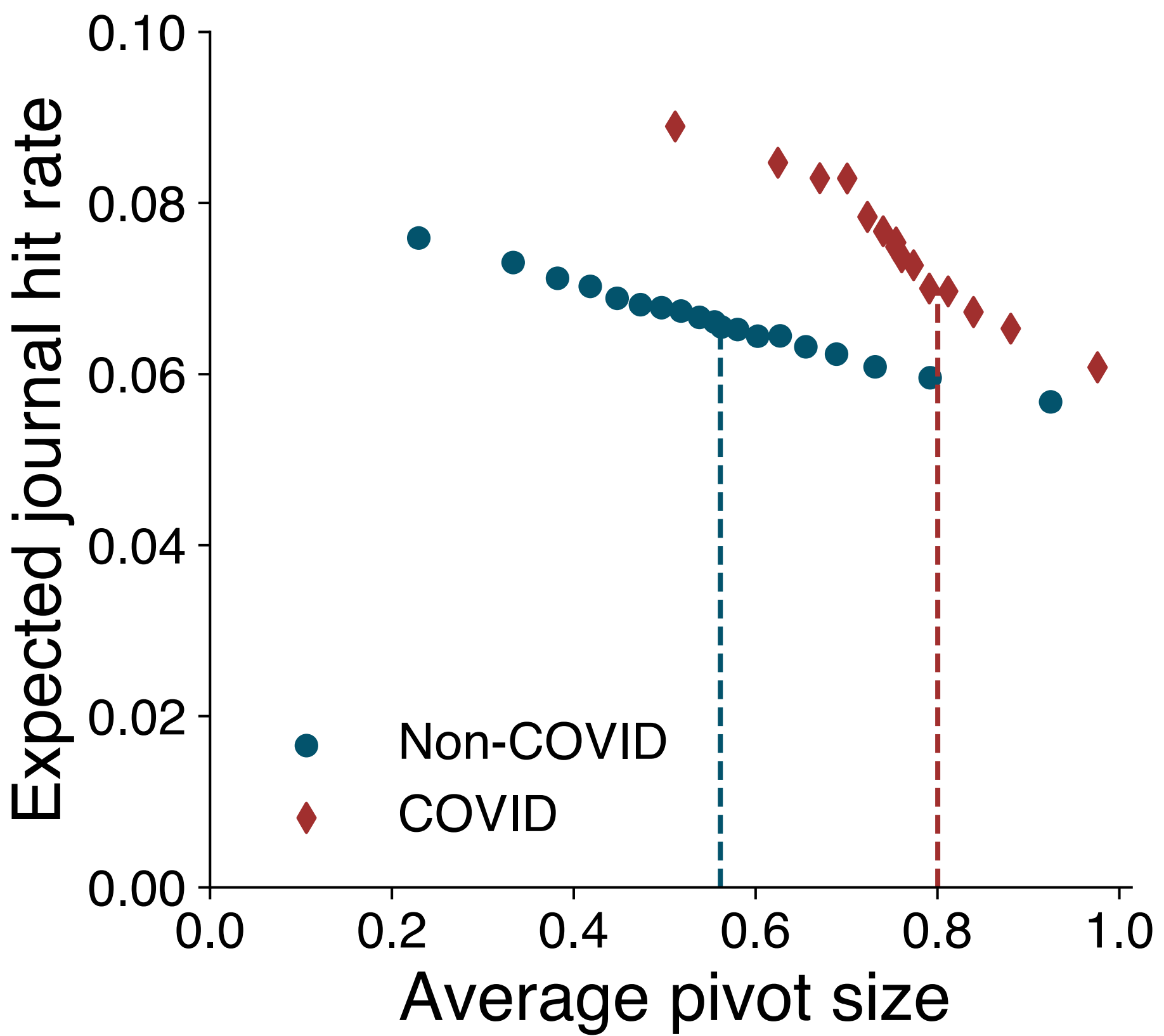

**Figure S18: Hit rates and pivot size using individual fixed effects.** This figure follows the pivot-impact analysis shown in Fig. 4. In this version, we use a regression adjustment for individual fixed effects within each series to control for unobservable factors that might drive pivot size and impact differentially across researchers.

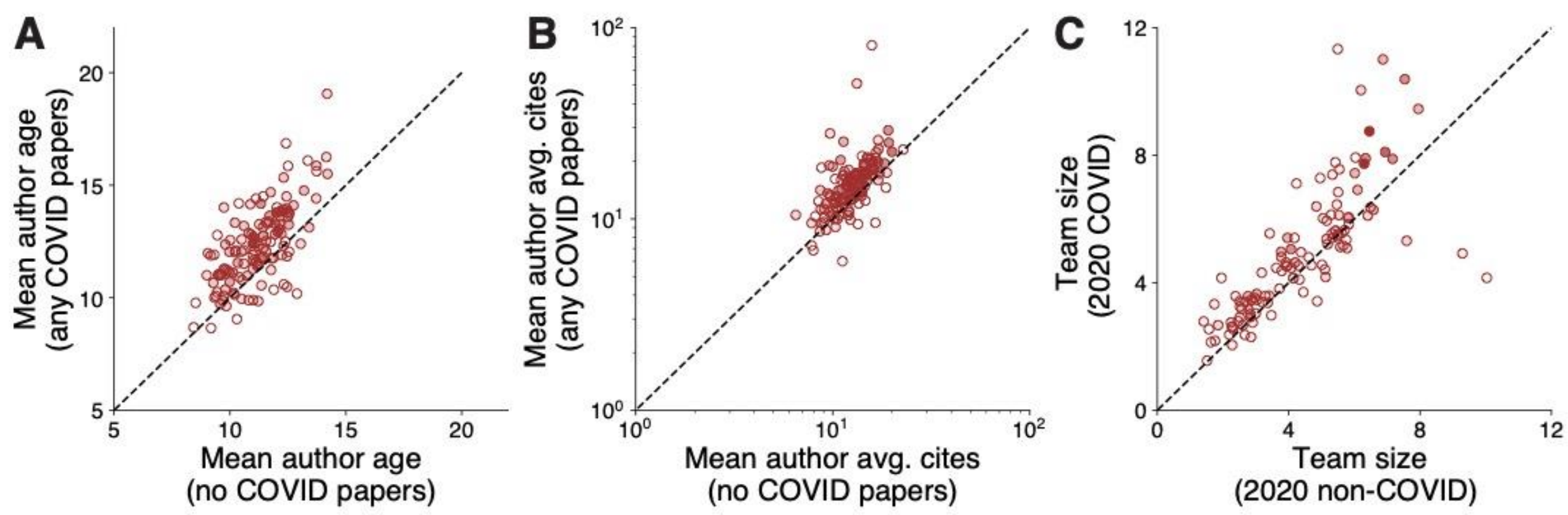

**Figure S19: Pivoting characteristics by field.** These plots examine paper and author features by field, comparing COVID and non-COVID research among actively publishing scientists in 2020. Markers with darker shading indicate fields with more COVID publications. Authors are assigned to the level-1 field in which they have published the most. **(A)** Mean author age for those who write COVID-19 papers is greater than for those who do not in 82% of fields. **(B)** Mean author prior impact for those who write COVID-19 papers is greater than those who do not in 83% of fields. **(C)** Mean team size is higher for COVID-19 papers in 77% of fields.

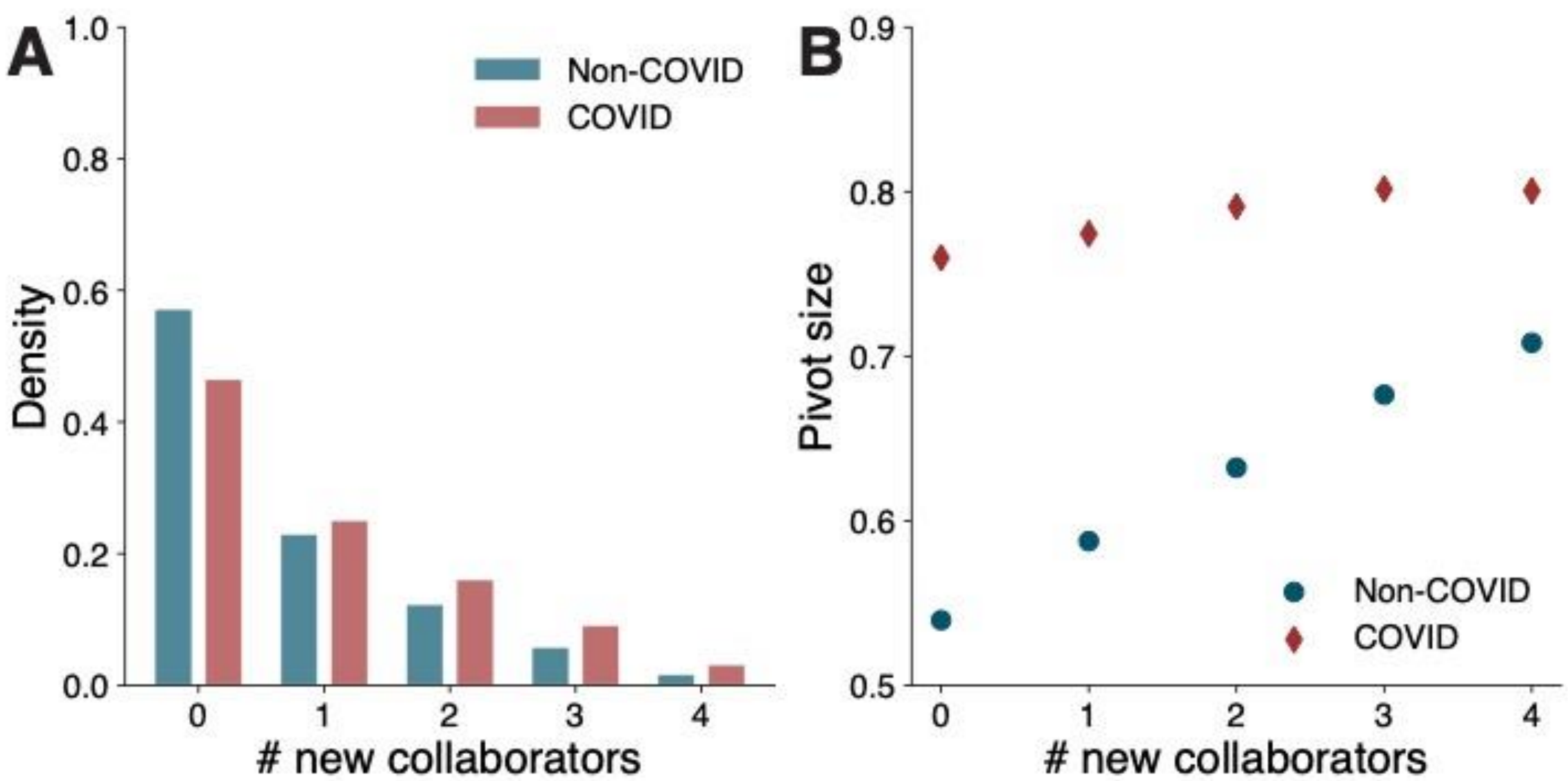

**Figure S20: Pivots and new collaborators**. These plots consider all 2020 publications with exactly five authors (similar results are found using different team sizes). **(A)** Papers with no new coauthors are the most common form, while **(B)** pivot size is increasing with the number of new coauthors.

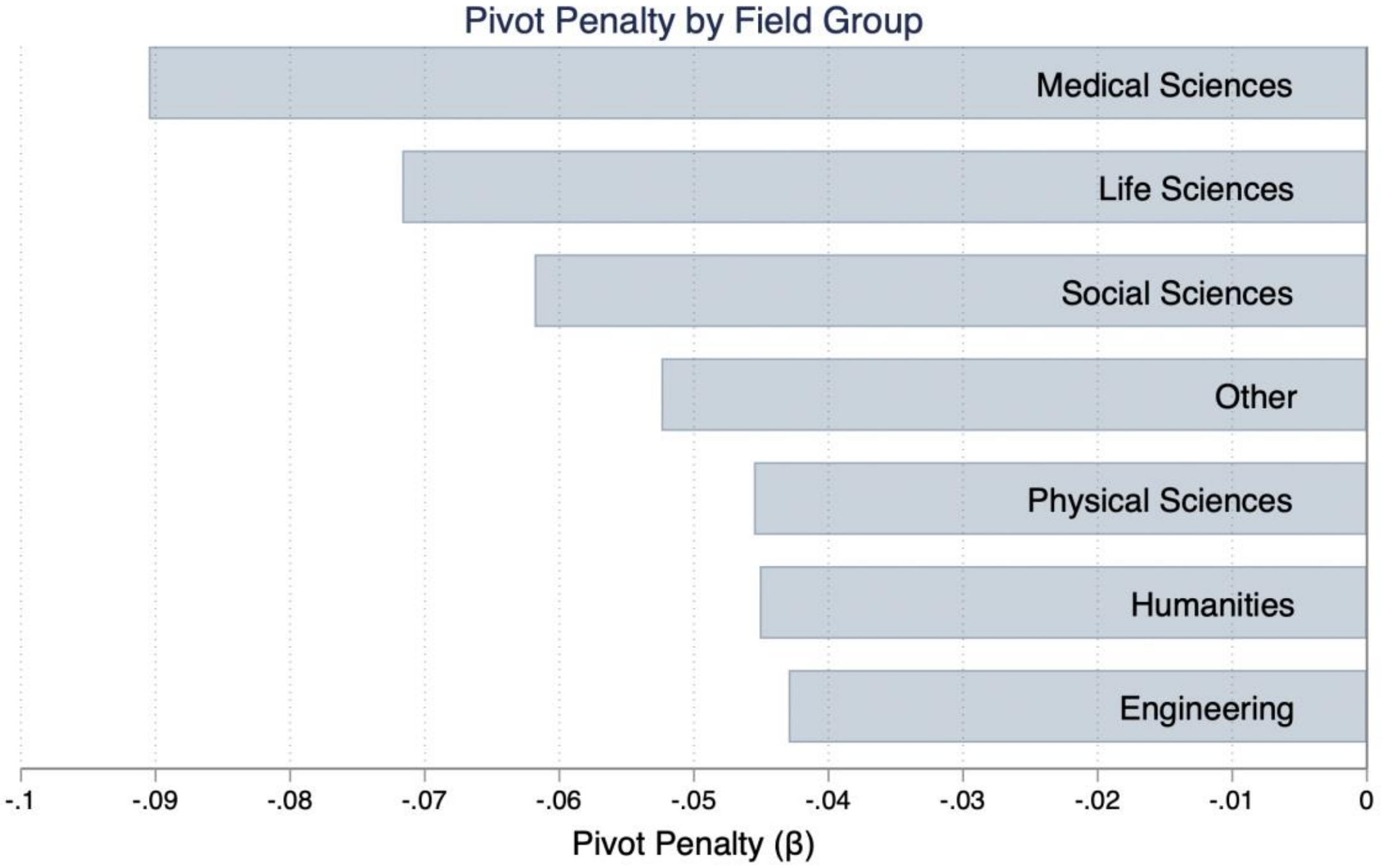

**Figure S21: Pivot penalties by field groups**. This figure explores heterogeneity in the pivot penalty according to higher level field groupings. Magnitudes vary across field groups, but the pivot penalty appears substantial in diverse areas of study. See Section S3.2 for further discussion.

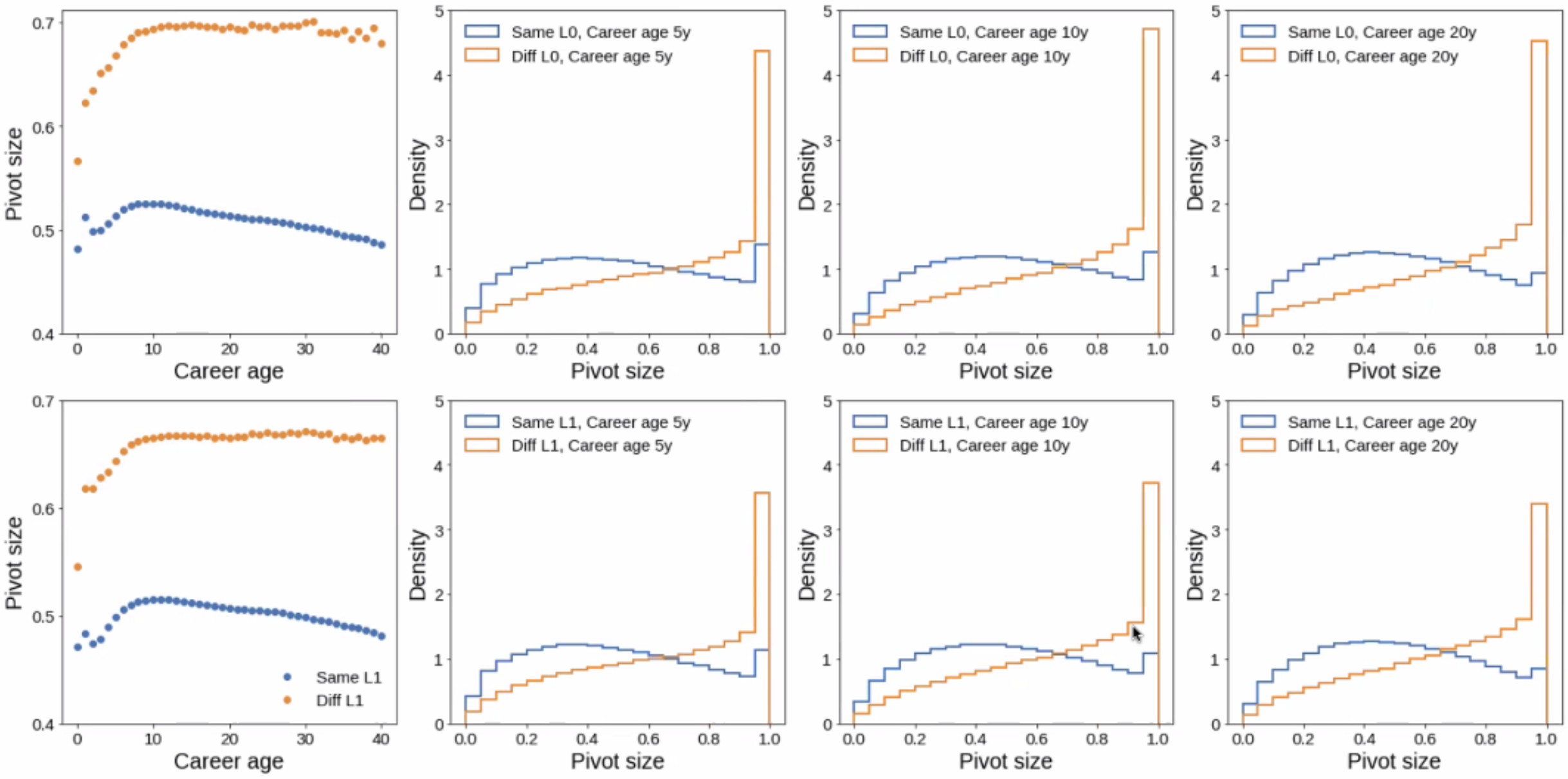

**Figure S22: Pivot size and field switching**. This figure considers the relationship between pivot size and binary indicators for a change in field. When researchers change field with a new paper, mean pivot sizes tend to be substantially larger, both when switching among the 22 Level-0 fields (**top left**) or switching among the 154 Level-1 fields (**bottom left**). Moreover, the binary measure of field switches is strongly associated with pivot sizes close to 1 (**other panels**). These findings are broadly similar at different stages of the career. See Section S3.3 for discussion.

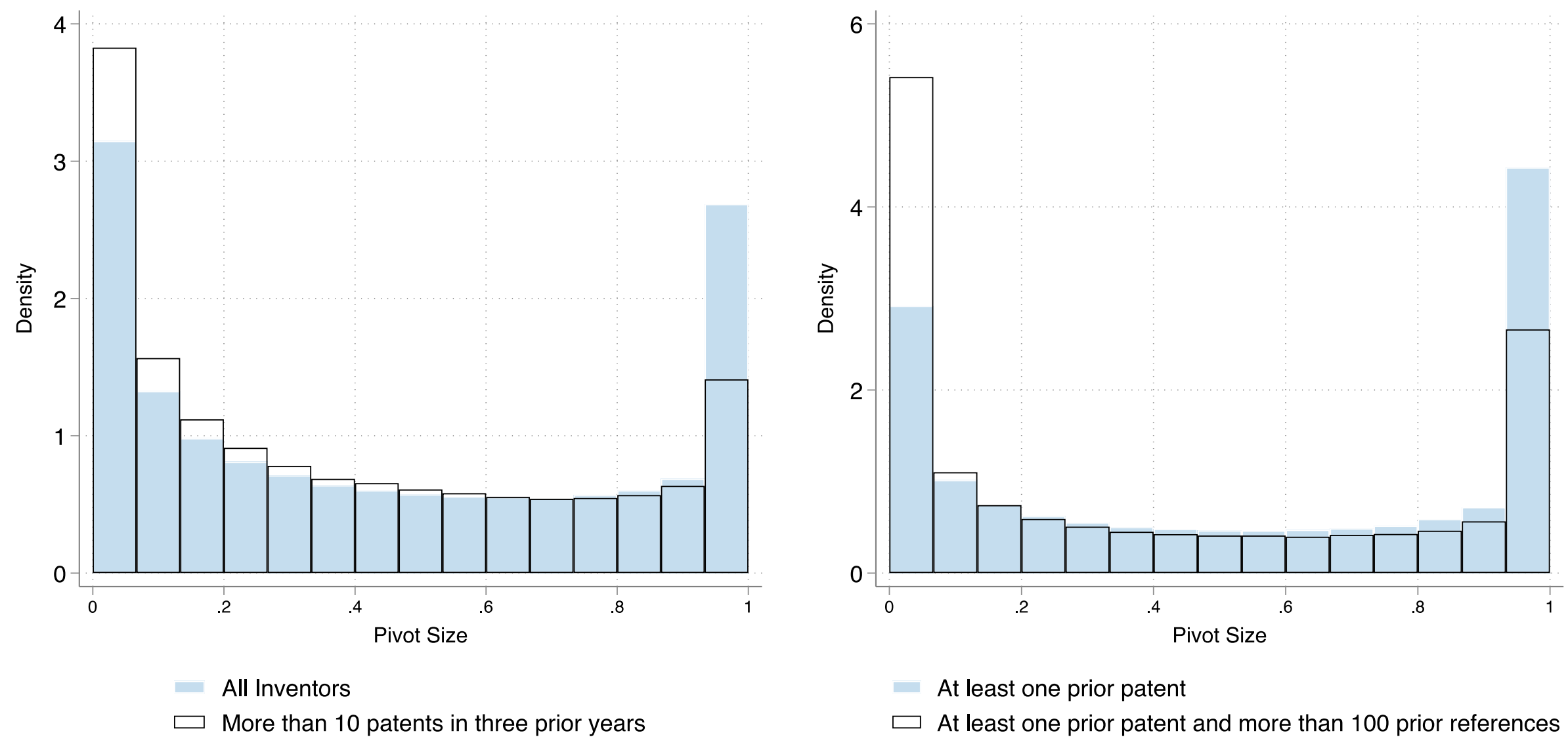

**Figure S23: Bimodal pivot distribution for patents**. This figure further explores the bimodal nature of the patent pivot size distribution. The pivot size distribution shifts leftward when restricting the sample to inventors with at least 10 patents in the prior three years (left panel). The pivot size distribution also shifts left when we restrict the sample to inventors with exactly one patent in the prior three years, but then separate out cases where that patent has at least 100 prior art references (right panel). While the presence of very high pivot patents declines substantially, the bimodal nature of the patent relationship remains. Thus the bimodal distribution of patents is not due to cases with a small set of reference material. See Section S2.2.2 for further discussion.

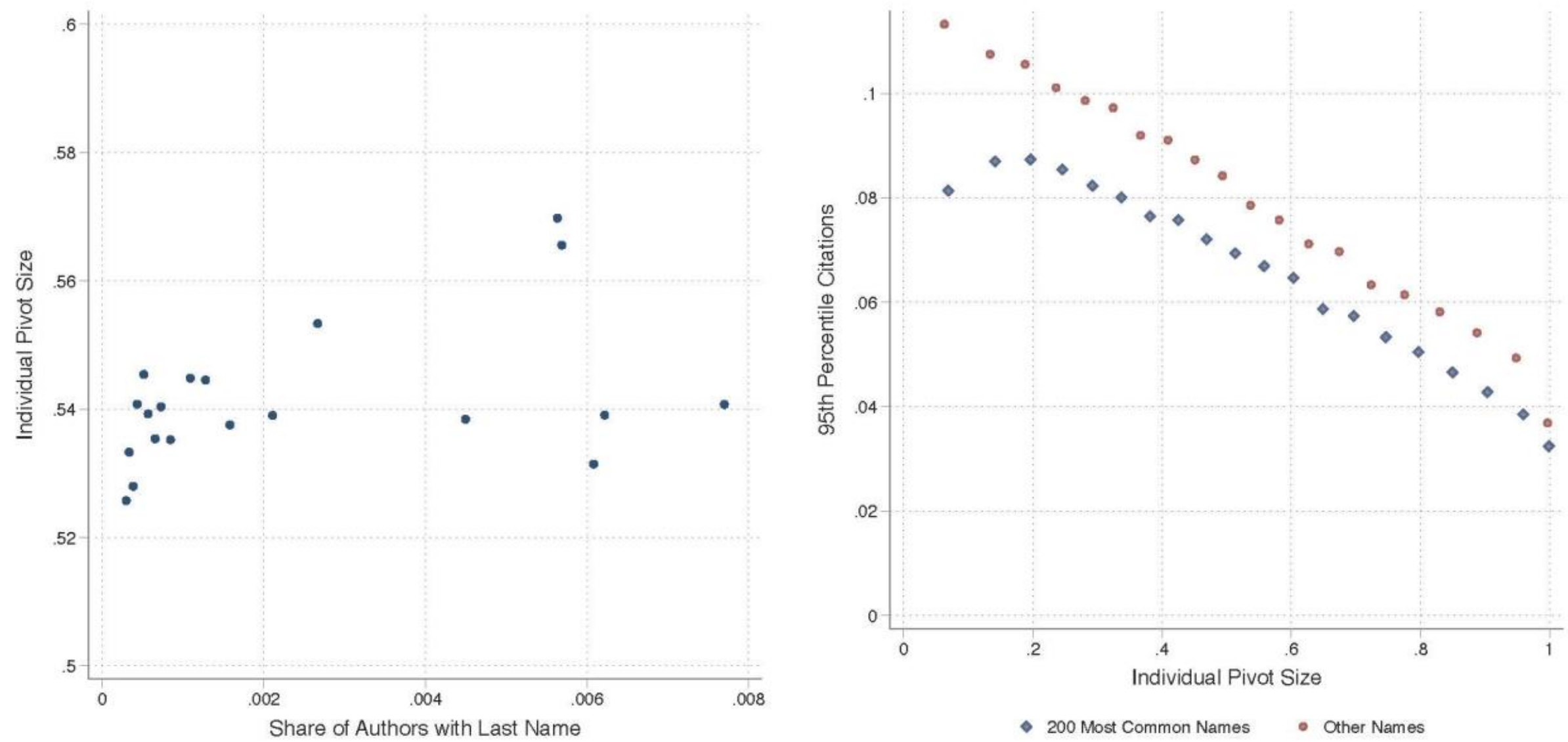

**Figure S24: Common and rare names**. These figures examine the findings in light of potential name disambiguation challenges. We compare results for scientists with common and rare names, where common names may be harder to disambiguate. (**A**) Mean pivot size as function of surname frequency. This plot shows a bin scatter and indicates little relationship between how common a name is and the mean pivot size for the individual. (B) The pivot penalty is robust when analyzed separately among scholars with common surnames or less common surnames. See Section S3.1 for discussion.

| Panel A | Share of L1 fields with negative correlation between pivot size and impact: | Number of fields with at least 20 papers |
|---|---|---|
| All 1970-2019 papers | 93.5% | 153 |
| All 2020 papers | 88.2% | 149 |
| Non-COVID 2020 papers | 89.5% | 149 |
| COVID 2020 papers | 59.5% | 111 |
| **Panel B** | **Share of L1 fields where correlation is becoming more negative over time:** | |
| All 1970-2019 papers | 88.2% | 153 |

**Table S1: Pivot-impact relationship by scientific field.** This table shows that a large majority of fields exhibit negative relationships between pivot size and impact. Further, this relationship is becoming more negative over time. In the 1970-2019 rows, impact is measured as an indicator for being in the 95$^{th}$ percentile of citations by year and field. In the 2020 rows, impact is measured as the journal hit rate, or the probability that a paper will reach the 95$^{th}$ percentile of citations based on journal placement. In all rows, only fields with at least 20 papers are included in the share, with the number of qualifying fields listed for each row. In Panel A, the sign of the relationship is estimated within each field using linear regression of impact regressed on pivot size. In Panel B, we add to the field-specific regressions an interaction between pivot size and year to estimate the change in slope over time.

| Panel A | Share of classes with negative correlation between pivot size and impact: | Number of classes with at least 20 patents |
|---|---|---|
| All 1980-2015 patents | 91.3% | 127 |
| Panel B | Share of classes where correlation is becoming more negative over time: | |
| All 1980-2015 patents | 76.4% | 127 |

**Table S2: Pivot-impact relationship by patent class.** This table shows that a large majority of patent classes exhibit negative relationships between pivot size and impact. Further, this relationship is becoming more negative over time. Impact is measured as an indicator for being in the 95$^{th}$ percentile of citations by year and field. In all rows, only classes with at least 20 patents are included in the share, with the number of qualifying fields listed for each row. In Panel A, the sign of the relationship is estimated within each field using linear regression of impact regressed on pivot size. In Panel B, we add to the field-specific regressions an interaction between pivot size and year to estimate the change in slope over time.

| | (1) | (2) | (3) | (4) | (5) |
|---|---|---|---|---|---|
| Pivot Size | -0.0687*** | -0.0706*** | -0.0714*** | -0.0688*** | -0.0699*** |
| | (0.000774) | (0.00115) | (0.00134) | (0.00186) | (0.00368) |
| Constant | 0.0906*** | 0.103*** | 0.109*** | 0.119*** | 0.151*** |
| | (0.000463) | (0.000625) | (0.000711) | (0.000939) | (0.00168) |
| Sample | At least 5 references | At least 15 references | At least 20 references | At least 30 references | At least 50 references |
| Observations | 1,337,008 | 890,402 | 737,279 | 472,913 | 175,472 |
| R-squared | 0.006 | 0.004 | 0.004 | 0.003 | 0.002 |

**Table S3: The pivot penalty for alternative thresholds for the number of cited references.** These analyses use publications in 2010. The dependent variable is an indicator for being in the upper 5$^{th}$ percentile of citations received for the field and publication year. Moving left to right, the columns increasingly restrict the sample, as indicated, according to the number of backwards references a paper makes. Robust standard errors in parentheses (*** p<0.01, ** p<0.05, * p<0.1).

| | (1) | (2) | (3) | (4) |
|---|---|---|---|---|
| Pivot Size | -0.0751*** | -0.0414*** | -0.0426*** | -0.0308*** |
| | (0.000521) | (0.000728) | (0.000788) | (0.000858) |
| Individual FE | | X | | |
| Individual X Field of Journal FE | | | X | |
| Individual X Journal FE | | | | X |
| Observations | 3,708,999 | 3,708,999 | 3,708,999 | 3,708,999 |
| R-squared | 0.005 | 0.215 | 0.299 | 0.453 |

**Table S4: The pivot penalty with individual, field, and journal fixed effects.** This table reports regressions of impact on pivot size. The dependent variable is an indicator for a paper reaching the 95th percentile of citations for the field and year. The regression sample is all papers published between 2005 and 2010 and where an author has multiple papers appear in the same field and journal. Individual fixed effects are added to the model in column 2. Individual by field of journal fixed effects are added in column 3, where the field of each journal is defined by the modal field of papers published in the journal. Individual by journal fixed effects are used in column 4. Heteroskedasticity-robust standard errors are reported in parentheses (*** p<0.01, ** p<0.05, * p<0.1).

| | (1) | (2) | (3) | (4) | (5) | (6) |
|---|---|---|---|---|---|---|
| Pivot Size | -0.0779*** | -0.0644*** | -0.0530*** | -0.0306*** | -0.0372*** | -0.0500*** |
| | (0.000772) | (0.00124) | (0.00125) | (0.00816) | (0.00874) | (0.0106) |
| Career Age x Pivot Size | | -0.00105*** | | | | |
| | | (8.28e-05) | | | | |
| Career Age | | 0.00102*** | | | | |
| | | (4.76e-05) | | | | |
| Constant | 0.116*** | 0.102*** | 0.103*** | 0.0795*** | 0.0883*** | 0.0888*** |
| | (0.000455) | (0.000747) | (0.000685) | (0.00454) | (0.00483) | (0.00575) |
| Individual Fixed Effects | No | No | Yes | Yes | Yes | Yes |
| Career Age Sample | All | All | All | 10+ years | 4-9 years | 1-3 years |
| Observations | 1,555,874 | 1,555,874 | 1,555,874 | 29,082 | 27,554 | 16,281 |
| R-squared | 0.007 | 0.007 | 0.369 | 0.404 | 0.405 | 0.412 |

**Table S5: The pivot penalty by career stage.** These analyses use publications in 2010. The dependent variable is an indicator for being in the upper 5$^{th}$ percentile of citations received for the field and publication year. Column (1) presents the baseline pivot penalty result with no controls. Column (2) shows a small negative interaction between pivot size and career stage in predicting impact. Researchers further in their career thus face somewhat large pivot penalties, although this steepening of the pivot penalty is small, and much smaller than the general pivot penalty. Column (3) presents a baseline specification with individual fixed effects. Columns (4)-(6) then run separate individual fixed effect regressions for the indicated range of career ages, further restricting the sample to authors with exactly the same publication rate over the prior three years (5 publications), to ensure similarity in productivity. These final analyses further show that pivot penalty appears at the earliest stages of the career. Robust standard errors in parentheses (*** p<0.01, ** p<0.05, * p<0.1).

**Panel A**

| | (1) | (2) | (3) | (4) | (5) | (6) | (7) |
|---|---|---|---|---|---|---|---|
| | | Reduced Form | | | | Two-Stage Least Squares | |
| | Pivot Size | Hit Paper | Hit Paper (2-yr) | Normalized Citations | Hit Paper | Hit Paper (2-yr) | Normalized Citations |
| Treated × post | 0.025*** | -0.004*** | -0.007*** | -0.041*** | | | |
| | (0.001) | (0.001) | (0.001) | (0.008) | | | |
| Post | -0.017*** | 0.000 | 0.001** | 0.007 | -0.003*** | -0.003*** | -0.021*** |
| | (0.001) | (0.001) | (0.001) | (0.007) | (0.000) | (0.000) | (0.006) |
| Pivot size | | | | | -0.164*** | -0.266*** | -1.642*** |
| | | | | | (0.024) | (0.024) | (0.313) |
| YearFE | X | X | X | X | X | X | X |
| AuthorFE | X | X | X | X | X | X | X |
| R-squared | 0.418 | 0.153 | 0.158 | 0.095 | - | - | - |
| Observations | 5,823,683 | 5,823,683 | 5,823,683 | 5,823,683 | 5,823,683 | 5,823,683 | 5,823,683 |

**Panel B**

| | (1) | (2) | (3) | (4) | (5) | (6) | (7) |
|---|---|---|---|---|---|---|---|
| | | Reduced Form | | | | Two-Stage Least Squares | |
| | Pivot Size | Hit Paper | Hit Paper (2-yr) | Normalized Citations | Hit Paper | Hit Paper (2-yr) | Normalized Citations |
| Treated × post | 0.037*** | -0.007*** | -0.011*** | -0.061*** | | | |
| | (0.001) | (0.001) | (0.001) | (0.015) | | | |
| Post | -0.012*** | -0.002*** | -0.001** | -0.014 | -0.004*** | -0.005*** | -0.033*** |
| | (0.001) | (0.001) | (0.001) | (0.008) | (0.001) | (0.001) | (0.009) |
| Pivot size | | | | | -0.191*** | -0.308*** | -1.668*** |
| | | | | | (0.035) | (0.036) | (0.411) |
| YearFE | X | X | X | X | X | X | X |
| AuthorFE | X | X | X | X | X | X | X |
| R-squared | 0.441 | 0.159 | 0.166 | 0.079 | - | - | - |
| Observations | 2,958,536 | 2,958,536 | 2,958,536 | 2,958,536 | 2,958,536 | 2,958,536 | 2,958,536 |

**Table S6: Difference-in-Differences analysis, retractions** This table presents the regression results for Figure 3 together with alternative specifications. The top table (panel A) considers the treatment sample defined by having cited a retracted paper at least once prior to its retraction. The bottom table (panel B) considers the treatment group defined as having cited a retracted

paper multiple times prior to its retraction. In both tables the columns are the same. Columns (1) considers the effect of the shock on pivot size. Columns (2)-(4) consider the reduced-form of the shock effect on impact. Impact is measured alternatively as the hit rate of the paper using the whole forward citation window after publication (2), the hit rate of the paper using only the first two years of citations after publication (3), and the citation count of the paper in ratio to the field-year mean (4). Columns (5)-(7) then consider these impact effects again using two-stage least squares. All regressions include individual fixed effects and year fixed effects. Standard errors are clustered at the author level and shown in parentheses (*** p<0.01, ** p<0.05, * p<0.1).

| | (1) | (2) | (3) | (4) | (5) | (6) | (7) |
|---|---|---|---|---|---|---|---|
| | Reduced Form | | | | Two-Stage Least Squares | | |
| | Pivot Size | Hit Paper | Hit Paper (2-yr) | Normalized Citations | Hit Paper | Hit Paper (2-yr) | Normalized Citations |
| Treated × post | 0.014*** | -0.015** | -0.019*** | -0.067** | | | |
| | (0.005) | (0.006) | (0.006) | (0.032) | | | |
| Pivot size | | | | | -1.117* | -1.355** | -4.868 |
| | | | | | (0.599) | (0.643) | (2.972) |
| YearFE | X | X | X | X | X | X | X |
| AuthorFE | X | X | X | X | X | X | X |
| R-squared | 0.309 | 0.096 | 0.086 | 0.085 | - | - | - |
| Observations | 56,257 | 56,257 | 56,257 | 56,257 | 56,257 | 56,257 | 56,257 |

**Table S7: Replication analysis** This table considers replication failures in psychology, where prior work researchers had been building upon is no longer seen as reliable. The table form follows the same structure as Table S6. See also Section S2.6 for discussion of methods. All regressions include individual fixed effects and year fixed effects. Standard errors are clustered at the author level and shown in parentheses (*** $p<0.01$, ** $p<0.05$, * $p<0.1$).

| Field | Number of Papers | Pivot-Impact Regression Estimates | | |
|---|---|---|---|---|
| | | Slope | Standard Error | P-value |
| Distributed Computing | 36295 | 0.012 | 0.005 | 0.017 |
| Computer Hardware | 15579 | 0.017 | 0.007 | 0.018 |
| Other Information and Computing Sciences | 11202 | 0.025 | 0.011 | 0.017 |
| Film, Television and Digital Media | 7088 | 0.018 | 0.016 | 0.259 |
| Other Earth Sciences | 3015 | 0.005 | 0.023 | 0.826 |
| Visual Arts and Crafts | 2195 | 0.122 | 0.024 | 0.000 |
| Art Theory and Criticism | 589 | 0.093 | 0.085 | 0.276 |
| Other Law and Legal Studies | 140 | 0.052 | 0.172 | 0.763 |
| Other Built Environment and Design | 87 | 0.450 | 0.187 | 0.018 |
| Other Philosophy and Religious Studies | 73 | 0.385 | 0.220 | 0.085 |

**Table S8: Outlier scientific fields** This table lists the 10 scientific fields that are outliers in showing a positive relationship between pivot size and impact. These 10 fields represent 6.5% of fields and only 0.18% of papers. See Section S3.2 for detailed discussion.

**Panel A**

| | (1) | (2) | (3) | (4) | (5) | (6) |
|---|---|---|---|---|---|---|
| Pivot Size | -0.101*** | | -0.0978*** | -0.104*** | | -0.0987*** |
| | (0.00172) | | (0.00175) | (0.00276) | | (0.00278) |
| Conventionality (median) | | 0.000143*** | 1.76e-05 | | 0.000130*** | -4.82e-06 |
| | | (1.22e-05) | (1.24e-05) | | (5.01e-05) | (5.00e-05) |
| Novelty (tail) | | -0.00118*** | -0.000971*** | | -0.00137*** | -0.00115*** |
| | | (3.26e-05) | (3.27e-05) | | (5.66e-05) | (5.67e-05) |
| Field FE | Y | Y | Y | Y | Y | Y |
| Conventionality Range | .25 to .75 percentile | .25 to .75 percentile | .25 to .75 percentile | .40 to .60 percentile | .40 to .60 percentile | .40 to .60 percentile |
| Observations | 379,346 | 379,346 | 379,346 | 151,646 | 151,646 | 151,646 |
| R-squared | 0.012 | 0.006 | 0.014 | 0.013 | 0.008 | 0.016 |

**Panel B**

| | (1) | (2) | (3) | (4) |
|---|---|---|---|---|
| Pivot Size | -0.108*** | -0.102*** | -0.106*** | -0.101*** |
| | (0.00391) | (0.00393) | (0.00872) | (0.00876) |
| | | 9.46e-05 | | 0.000568 |
| Conventionality (median) | | (0.000141) | | (0.00155) |
| | | -0.00112*** | | -0.000980*** |
| Novelty (tail) | | (8.18e-05) | | (0.000181) |
| Field FE | Y | Y | Y | Y |
| Conventionality Range | .45 to .55 percentile | .45 to .55 percentile | .49 to .51 percentile | .49 to .51 percentile |
| Observations | 75,868 | 75,868 | 15,212 | 15,212 |
| R-squared | 0.015 | 0.017 | 0.018 | 0.020 |

**Table S9: Pivots and combinations** This table presents regression evidence considering pivoting behavior, novel combinations, and conventional combinations together. The dependent variable is an indicator for being a high impact paper, defined as in the upper 5th percentile of citations received for the field and year. The analysis uses papers published in 2010. See Section S2.3 for details on the combinatorial measures. To test whether pivot size predicts citation impact net of conventionality and novelty, we control for all three variables independently and over narrow ranges of the conventionality score. In Panel A, the first three columns consider the middle 50 percent of observations by conventionality, while the second three columns consider the middle 20 percent of observations by conventionality. In Panel B, the first two columns consider the middle 10 percent of observations by conventionality, while the next two columns consider the middle 2 percent of observations by conventionality. These regressions continue to show a substantial pivot penalty. Standard errors in parentheses (*** $p<0.01$, ** $p<0.05$, * $p<0.1$).